\documentclass[fleqn,usenatbib]{mnras}

\usepackage{newtxtext,newtxmath}
\usepackage[T1]{fontenc}
\usepackage{ae,aecompl}

\usepackage{graphicx}	% Including figure files
\usepackage{amsmath}	% Advanced maths commands
\usepackage{multicol}
\usepackage[flushleft]{threeparttable}

\usepackage{amsmath} % or simply amstext

\usepackage{xcolor}

\newcommand{\cha}{{\it Chandra}}

\newcommand{\erg}{{\hbox{erg~s$^{-1}$}}}             

\newcommand{\cmsq}{cm$^{-2}$ }    
\newcommand{\nh}{$N_{\rm{H}}$}

\newcommand{\etal}{{{et~al.}}}

\newcommand{\msunpyr}{{\hbox{M$_{\odot}\,\rm{yr^{-1}}$}}}

\newcommand{\Ha}{H$\alpha$}
\newcommand{\Hb}{H$\beta$}

\def\x2{$\chi^{2}$}
\usepackage{tikz,xcolor}
\definecolor{lime}{HTML}{A6CE39}
\DeclareRobustCommand{\orcidicon}{%
	\begin{tikzpicture}
	\draw[lime, fill=lime] (0,0) 
	circle [radius=0.16] 
	node[white] {{\fontfamily{qag}\selectfont \tiny ID}};
	\draw[white, fill=white] (-0.0625,0.095) 
	circle [radius=0.007];
	\end{tikzpicture}
	\hspace{-2mm}
}
\foreach \x in {A, ..., Z}{%
	\expandafter\xdef\csname orcid\x\endcsname{\noexpand\href{https://orcid.org/\csname orcidauthor\x\endcsname}{\noexpand\orcidicon}}
}

\title[Metallicity-$L_X$ variations in NGC\,922]{Metallicity and X-ray luminosity variations in NGC\,922}

\author[K. Kouroumpatzakis \etal]{K. Kouroumpatzakis$^{1,2}$\orcidA{}\thanks{E-mail: kkouroub@physics.uoc.gr},
\author x A. Zezas$^{1,2,3}$\orcidB{},
\author x A. Wolter$^{4}$\orcidC{},
\author x A. Fruscione$^{3}$,
\newauthor
\author x K. Anastasopoulou$^{1,2}$\orcidE{},
\author [K. Kouroumpatzakis \etal] x and A. Prestwich$^{3}$\\
$^{1}$Department of Physics, Univ. of Crete, GR-70013 Heraklion, Greece\\
$^{2}$Institute of Astrophysics, FORTH, GR-71110 Heraklion, Greece\\
$^{3}$Center for Astrophysics \textbar\ Harvard \& Smithsonian\\
$^{4}$INAF-Osservatorio Astronomico di Brera, via Brera 28, I-20121 Milano, Italy\\
}

\date{Accepted XXX. Received YYY; in original form ZZZ}

\pubyear{2020}

\begin{document}
\label{firstpage}
\pagerange{\pageref{firstpage}--\pageref{lastpage}}
\maketitle

% Abstract of the paper
\begin{abstract}
We present a systematic study of the metallicity variations within the collisional ring galaxy NGC\,922 based on long-slit optical spectroscopic observations. 
We find a metallicity difference between star-forming regions in the bulge and the ring, with metallicities ranging from almost solar to significantly sub-solar ($\rm{[12+\log(O/H)]\sim 8.2}$). 
We detect \ion{He}{I} emission in the bulge and the ring star-forming regions indicating ionization from massive stars associated with recent ($<10$\,Myr) star-formation, in agreement with the presence of very young star-clusters in all studied regions.
We find an anti-correlation between the X-ray luminosity and metallicity of the sub-galactic regions of NGC\,922. 
The different regions have similar stellar population ages leaving metallicity as the main driver of the anti-correlation.
The dependence of the X-ray emission of the different regions in NGC\,922 on metallicity is in agreement with similar studies of the integrated X-ray output of galaxies and predictions from X-ray binary population models.
\end{abstract}

% Select between one and six entries from the list of approved keywords.
% Don't make up new ones.

%%%%%%%%%%%%%%%%%%%%%%%%%%%%%%%%%%%%%%%%%%%%%%%%%%
%%%%%%%%%%%%%%%%% BODY OF PAPER %%%%%%%%%%%%%%%%%%
%%%%%%%%%%%%%%%%%%%%%%%%%%%%%%%%%%%%%%%%%%%%%%%%%%

\section{Introduction}
\label{sec:Introduction}

Ring galaxies (RiGs) appear to form after a very special interaction where a small companion passes through a disk galaxy \citep[e.g.][]{1976ApJ...209..382L,1976ApJ...208..650T,1993MNRAS.261..804H,1994ApJ...437..611M}. 
This gravitational perturbation generates symmetrical waves or caustics through the galactic disk \citep[e.g.][]{1990ApJ...358...99S} leading to the creation of an enhanced star-formation ring \citep[e.g.][]{1997AJ....113..201A}. 
This relatively brief interaction gives rise to coeval star-formation in the ring, which 
provides an excellent environment to study the star-forming activity, neutral gas distribution, and metallic abundance in galaxy interactions.

The gas-phase metallicity is a key characteristic of any galactic environment.
For example there is a well-known correlation between metallicity and stellar mass ($M_\star$) in the general galaxy population  \citep[e.g.][]{1979A&A....80..155L,2004ApJ...613..898T,2008ApJ...681.1183K}.
Studies of local star-forming galaxies have shown a negative metallicity gradient with increasing galactocentric radius \citep[e.g.][]{1992MNRAS.259..121V,1994ApJ...420...87Z,1998AJ....116.2805V,2010ApJS..190..233M,2018MNRAS.476.3883L}.
RiGs provide a unique environment to explore the effect of a quasi-symmetric radial disturbance of the galaxy disk on its metallicity gradient. 
Prominent RiGs (e.g. the Cartwheel galaxy, Arp\,147, Lindsay-Shapley ring, Arp 284) show overall sub-solar metallicities \citep[e.g.][]{1977MNRAS.178..473F,1982MNRAS.199..633F,1997ApJ...474..686H,1997ApJ...478..112G,2011MNRAS.417..835F}.
However, there are cases of rings in RiGs with higher oxygen and nitrogen abundance compared to their bulges
\citep[e.g.][]{1998AJ....116.2757B,2019MNRAS.486.4186E}. 
This indicates that overall RiGs do not follow the metallicity gradient profile seen in disk galaxies.
This could be because of the mixing of gas from different regions of the disk as a result of the interaction.

An interesting feature of RiGs, that is directly linked to the age of the stellar populations in the ring and/or their metallicity, is their association with populations of Ultra-Luminous X-ray sources (ULXs).
These are generally defined as X-ray sources with luminosity in excess of $10^{39}$\erg \citep[e.g.][and references therein]{2017ARA&A..55..303K}.
RiGs show an excess of X-ray luminosity ($L_X$) and number of ULXs compared to typical star-forming galaxies \citep[e.g.][]{1999A&A...342...41W,2015MNRAS.448..781W}.
For example, the Cartwheel galaxy, the epitome of nearby RiGs, shows the largest number of ULXs (16) for a single galaxy \citep{2004A&A...426..787W}.  
In addition, the X-ray luminosity function (XLF) of RiGs appears to be flatter than the typical XLF of star-forming galaxies, although with the current data this difference is not statistically significant \citep[e.g.][]{2018ApJ...863...43W}.

It has been proposed that the observed excess of ULXs in RiGs is driven by the low metallicity of their galactic environment \citep[e.g.][]{2009MNRAS.395L..71M}.
Indeed, more recent studies support the idea that the 
X-ray luminosity per unit star-formation rate ($L_X$/SFR) is a function of metallicity, favoring low metallicity environments \citep[e.g.][]{2016ApJ...827L..21F,2016MNRAS.457.4081B,2017ApJ...840...39M,2020MNRAS.495..771F}.

NGC\,922 is a C-shaped galaxy with an off-centre star-forming bar and a semi-complete star-forming ring that is the result of an off-axis passage of a dwarf companion through the disk of a spiral galaxy \citep{2006MNRAS.370.1607W}. 
It has a recession velocity $v_{r}=3082.46\pm{5.40}$\,km/s corresponding to a distance of $42.46 \pm{2.48}$\,Mpc \citep{2004AJ....128...16K}.
It contains a higher abundance of neutral gas for a galaxy of its size, compared to typical star-forming galaxies \citep{2018MNRAS.476.5681E}. 
The interaction with the dwarf companion has triggered a star-formation episode in the bulge $\sim300$\,Myr ago, that continues until now. 
The ring on the other hand is dominated by very recent star-forming activity ($<10$\,Myr ago), as witnessed by a population of very young star clusters ($\sim 7$\,Myr), while in the bulge there is a combination of young and older star clusters \citep[$\gtrsim 100$\,Myr;][]{2010AJ....139.1369P}.

NGC\,922 hosts a population of many bright X-ray sources, including nine ULXs which is at odds with its near-solar metallicity \citep{2012ApJ...747..150P}. 
\cite{2012ApJ...747..150P} find that the number of ULXs per SFR in NGC\,922 is higher than that of the Cartwheel galaxy (but consistent within the uncertainties), despite the near-solar metallicity reported for NGC\,922. 
Furthermore, this ULX rate is higher than the average ULX/SFR rate found for late-type galaxies, but consistent with that found for Sc/Sm or irregular galaxies \citep{2020MNRAS.498.4790K}.
All these characteristics make NGC\,922 a perfect candidate to study the metallicity variations in RiGs, their relation to the dynamics of the interaction, and their effect on the luminous XRB populations.

In this paper we present long-slit observations of NGC\,922. 
We extracted optical spectra and measured emission-line fluxes and gas-phase metallicities from regions spread on the disk of NGC\,922 covering the bulge, the ring, and intermediate positions. 
We correlate these metallicity measurements with spatially resolved measurements of the stellar mass, SFR, and X-ray luminosity, based on archival multi-wavelength data.  
In Sections 2 and 3 we present the details of the optical spectroscopic and X-ray  observations and the data analysis respectively. 
The results of the analysis are presented in Section 4.
In Sections 5 and 6 we discuss and summarize the results.

In the following analysis we assume a cosmology with $\rm \Omega_m=0.3$, $\Omega_\Lambda=0.7$, $h=0.7$, and distance $D=42.46$ Mpc. 
We adopt as solar abundances $\rm Z_{\odot} = 0.0142$, $\rm X_{\odot} = 0.7154$, and [12 + log(O/H)$_\odot$] = 8.69 from \cite{doi:10.1146/annurev.astro.46.060407.145222}.

\section{Observations}
\label{sec:Observations}

We acquired long slit observations targeting the NGC\,922 galaxy with the European Southern Observatory (ESO) 3.58m New Technology Telescope (NTT) through the ESO Faint Object Spectrograph and Camera (EFOSC2\footnote{\url{https://www.eso.org/sci/facilities/lasilla/instruments/efosc.html}}). 
We obtained spectra for three different slit positions  on the galaxy.
We used the 1.5"-wide slit for slit rotations 1 ($\rm P.A.=68.7^{\circ}$) and 2 ($\rm P.A.=-39.7^{\circ}$), and the 2"-wide slit for slit rotation 3 ($\rm P.A.=-61.1^{\circ}$). 
The slit widths correspond to 309 and 412 pc respectively at the distance of the galaxy.
The apertures of the spectra extractions were even wider, enough to include emission from multiple star clusters, and limit the stochasticity effects.
We used two EFOSC2 grisms: (a) grism \#11 which covers the wavelength range of $\lambda$\,3403--7493\,\AA, at a dispersion of 4.1\,\AA/pixel and provides a resolution of  17.2--19.0\,\AA\,(FWHM)  at 3727 and 7136\,\AA\ respectively (\textit{low-resolution} grism), and (b) grism \#18 which covers the wavelength range $\lambda$\,4761--6754\,\AA, at a dispersion of 2.0\,\AA/pixel, and provides a resolution of 6.7--8.6\,\AA\,(FWHM)\footnote{The reported wavelength resolutions are for the 1.5"-wide slit.} at 5007 and 6563\,\AA\ respectively (\textit{high-resolution grism}).
The observations were performed under photometric conditions.
Observations with slit rotation 1 and 2 received a total exposure of 5400 seconds split in 6 frames. 
The spectrum obtained at slit rotation 3 had a total exposure of 1800 seconds, split in three frames.

\begin{figure*}
\begin{center}
\includegraphics[width=1.9\columnwidth]{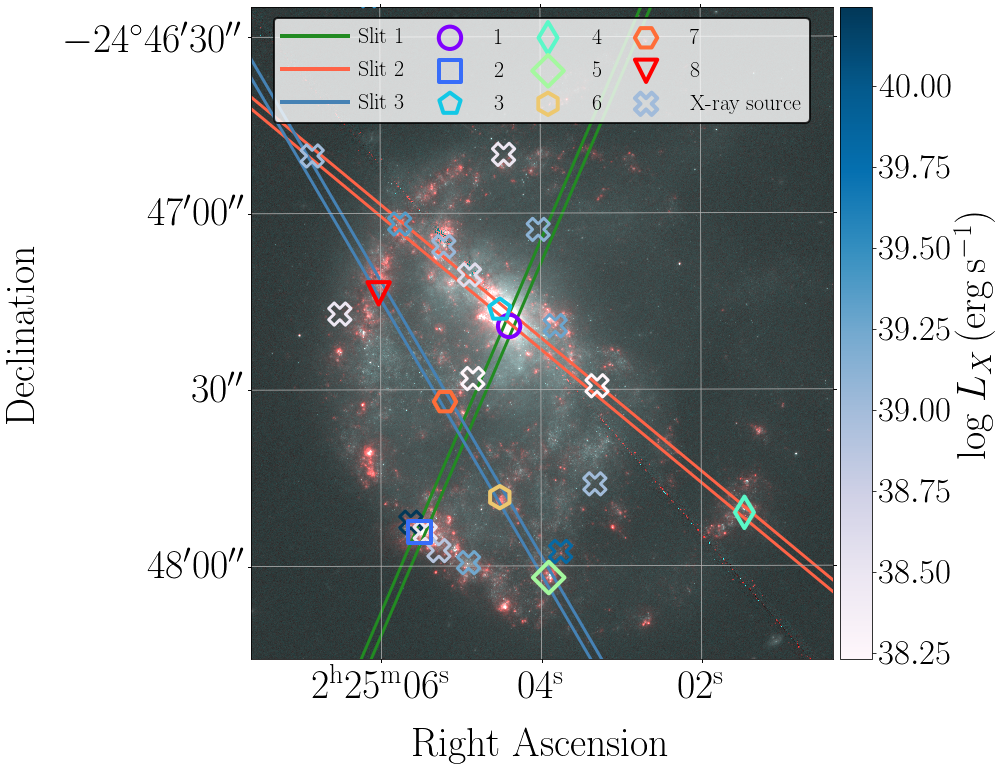}
\caption{
Colour composite image of NGC\,922 based on HST-WFC3 \Ha\ (F665N; red) and red continuum (F621M; blue/green). 
The positions of the slits used for this work, and the locations of the individual regions we extracted spectra from, are shown with unique colour and marker style, following the region IDs listed in Table \ref{tab:Obs}. 
X-ray point sources are presented with an open \textbf{x} symbol. They are colour-coded according to their X-ray luminosity (calculated from col. \texttt{flux\_aper\_b} in the CSC).
}
    \label{fig:Slit_positions}
\end{center}
\end{figure*}

We extracted spectra for a total of eight regions that show significant emission in the two-dimensional spectra. The coordinates and the observational parameters for each extraction region are given in Table \ref{tab:Obs}. 
Regions 1, 2, 3, 4 were  observed with both high and low resolution (grisms \#11, \#18). 
Regions 5, 6, 7, and 8 were observed only with the low resolution grism \#11. 
Therefore a total of 12 spectral extractions were analyzed, listed in Table\,\ref{tab:Obs}.
The lowest signal-to-noise ratio (S/N) spectrum extraction has a median $\rm S/N \sim 4.0$, at the continuum, with the emission lines having significantly higher S/N.
In Figure\,\ref{fig:Slit_positions} we present the slit placements along with the regions with spectra extractions (following the numbering convention of Table \ref{tab:Obs}), overlaid on a composite \Ha\ (F665N) and red continuum (F621M) image of the galaxy from archival HST-WFC3 data (P.I.: A. Prestwich; Program 11836). 
We also show the location of the X-ray sources obtained from the \textit{Chandra Source Catalog}\footnote{\url{https://cxc.harvard.edu/csc/}} \citep[CSC;][]{2010ApJS..189...37E}. 
The X-ray sources in Figure\,\ref{fig:Slit_positions} are colour-coded according to their X-ray luminosity (derived from the flux in the CSC \texttt{flux\_aper\_b} and the distance of the galaxy). 

\begin{table*}
    \centering
    \caption{Long slit observations and spectral extractions summary.}
    \begin{tabular}{c|c|c|c|c|c|c}
    \setlength\tabcolsep{0.pt}
         Region & Slit rotation & slit pos. angle & Resolution & frames $\times$ exp. time & R.A. & Dec. \\
         ID \#& ID \# & degrees & grism \# & \# $\times$ (sec) & (J2000) & (J2000)\\
         \hline
         1 & 1 & $68.7^{\circ}$ & 11 & $3 \times 900$ & 02:25:04.4 & -24:47:19.2\\
         1 & 1 & $68.7^{\circ}$ & 18 & $3 \times 900$ & 02:25:04.4 & -24:47:19.2\\
         2 & 1 & $68.7^{\circ}$ & 11 & $3 \times 900$ & 02:25:05.5 & -24:47:54.2\\
         2 & 1 & $68.7^{\circ}$ & 18 & $3 \times 900$ & 02:25:05.5 & -24:47:54.2\\
         3 & 2 & $-39.7^{\circ}$ & 11 & $3 \times 900$ & 02:25:04.5 & -24:47:16.4\\
         3 & 2 & $-39.7^{\circ}$ & 18 & $3 \times 900$ & 02:25:04.5 & -24:47:16.4\\
         4 & 2 & $-39.7^{\circ}$ & 11 & $3 \times 900$ & 02:25:01.5 & -24:47:51.0\\
         4 & 2 & $-39.7^{\circ}$ & 18 & $3 \times 900$ & 02:25:01.5 & -24:47:51.0\\
         5 & 3 & $-61.1^{\circ}$ & 11 & $3 \times 600$ & 02:25:03.9 & -24:48:02.0\\
         6 & 3 & $-61.1^{\circ}$ & 11 & $3 \times 600$ & 02:25:04.5 & -24:47:48.3\\
         7 & 3 & $-61.1^{\circ}$ & 11 & $3 \times 600$ & 02:25:05.2 & -24:47:32.0\\
         8 & 3 & $-61.1^{\circ}$ & 11 & $3 \times 600$ & 02:25:06.0 & -24:47:13.6\\
    \end{tabular}
    \label{tab:Obs}
\end{table*}

\section{Data analysis}
\label{sec:Analysis}

\subsection{Optical spectra}
\label{sec:optical_spectra}

We first performed the  basic reductions, such as bias subtraction and flat fielding for all the observed images. 
Because we are interested in spectra from different regions along the slits, the wavelength calibration was performed on the two-dimensional spectra in order to correct for slit distortions. 
We used spectra from HeNeAr lamps obtained before and after each observation. 
We used the \texttt{IRAF} \citep{1986SPIE..627..733T,1993ASPC...52..173T} tasks \texttt{identify} and \texttt{reidentify} in order to obtain wavelength calibrations for different locations along the slit, and the \texttt{transform} task in order to calculate the mapping from (x,y) pixel coordinates to ($\lambda$,y) coordinates.
We extracted the spectra from the combined images for each slit position using the \texttt{IRAF} \texttt{apall} task.  

The spectra were photometrically calibrated with observations of several spectrophotometric standard stars from the catalogue of \cite{1988ApJ...328..315M} obtained with each different instrumental setup.
The standard-star spectra were reduced the same way as the object spectra. 
The calibration (sensitivity function) was applied to the object spectra using the \texttt{IRAF} \texttt{sensfunc} task.

\begin{figure*} 
\begin{center}
\hbox{
\includegraphics[width=\columnwidth]{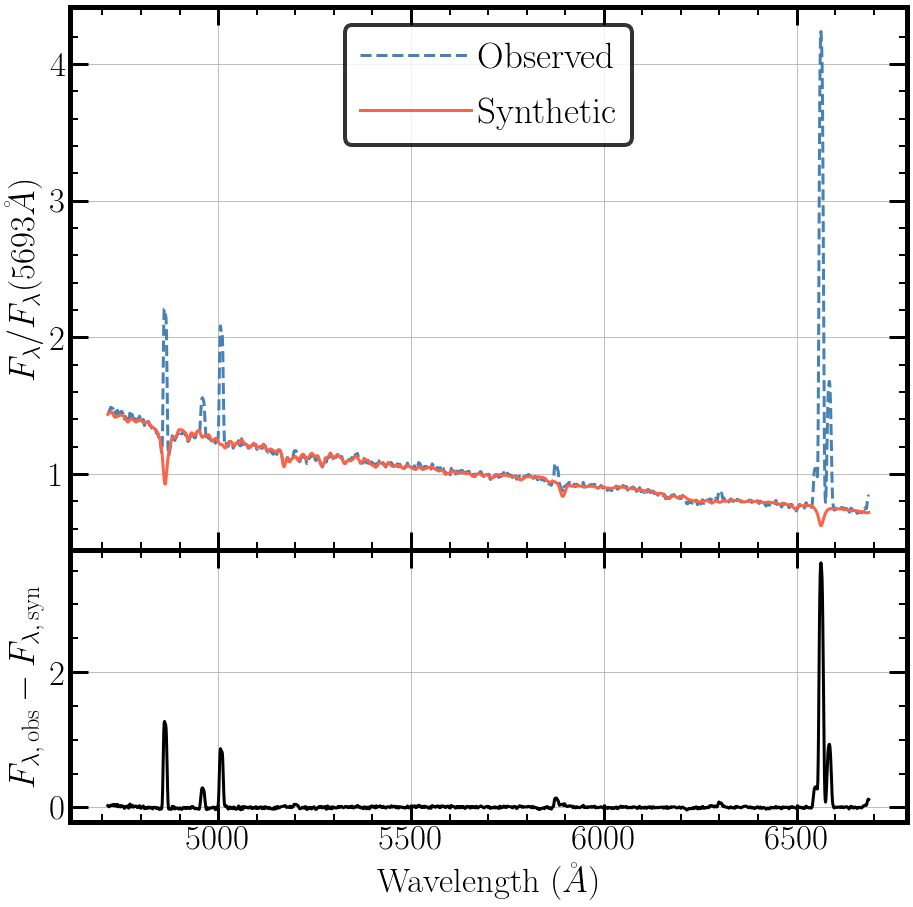}
\hspace{0.7truecm}
\includegraphics[width=\columnwidth]{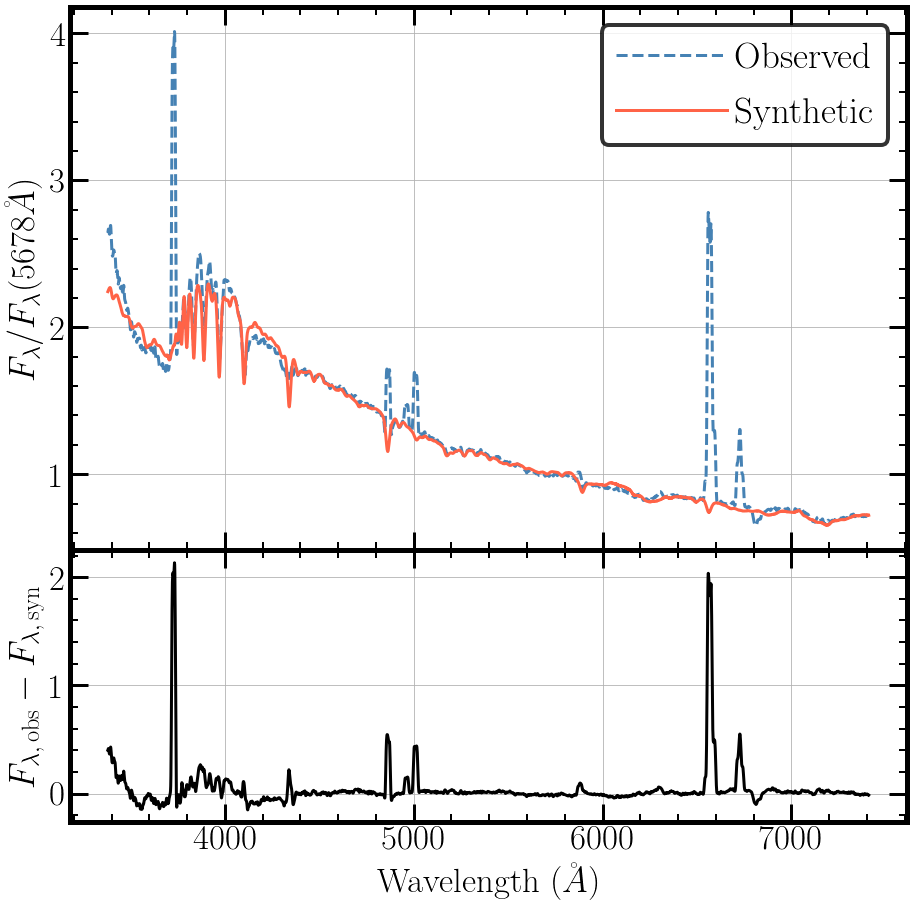}
}
\hbox{
\includegraphics[width=\columnwidth]{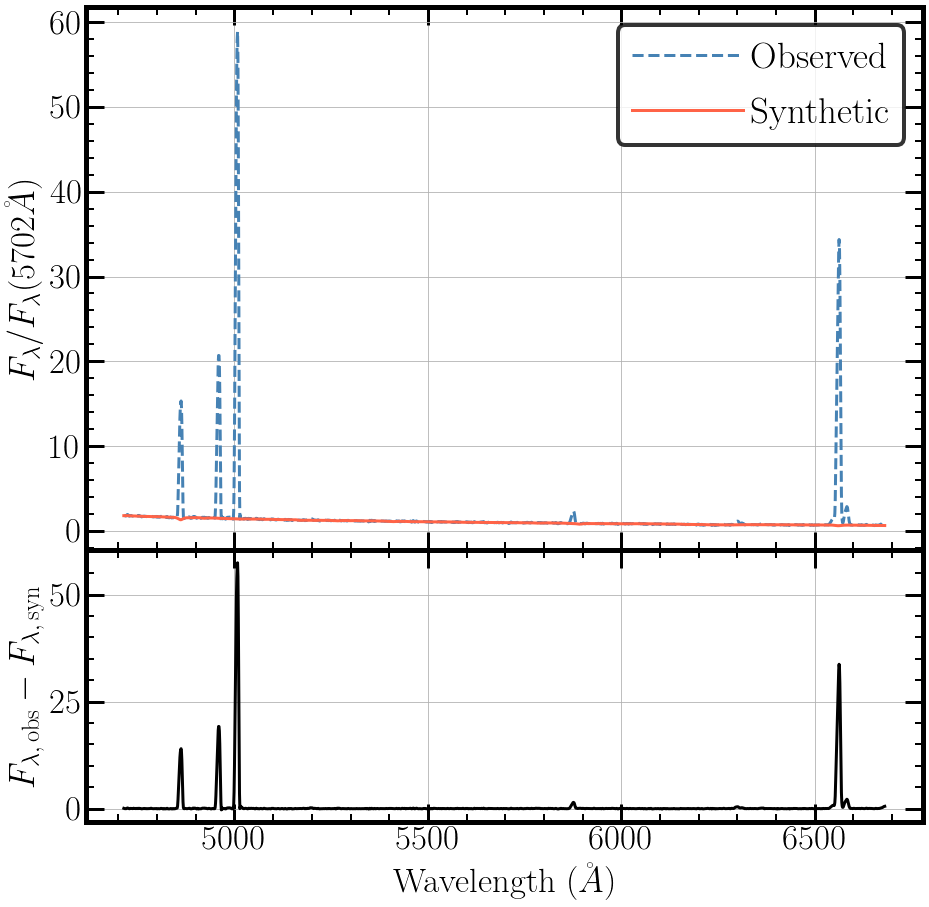}
\hspace{0.7truecm}
\includegraphics[width=\columnwidth]{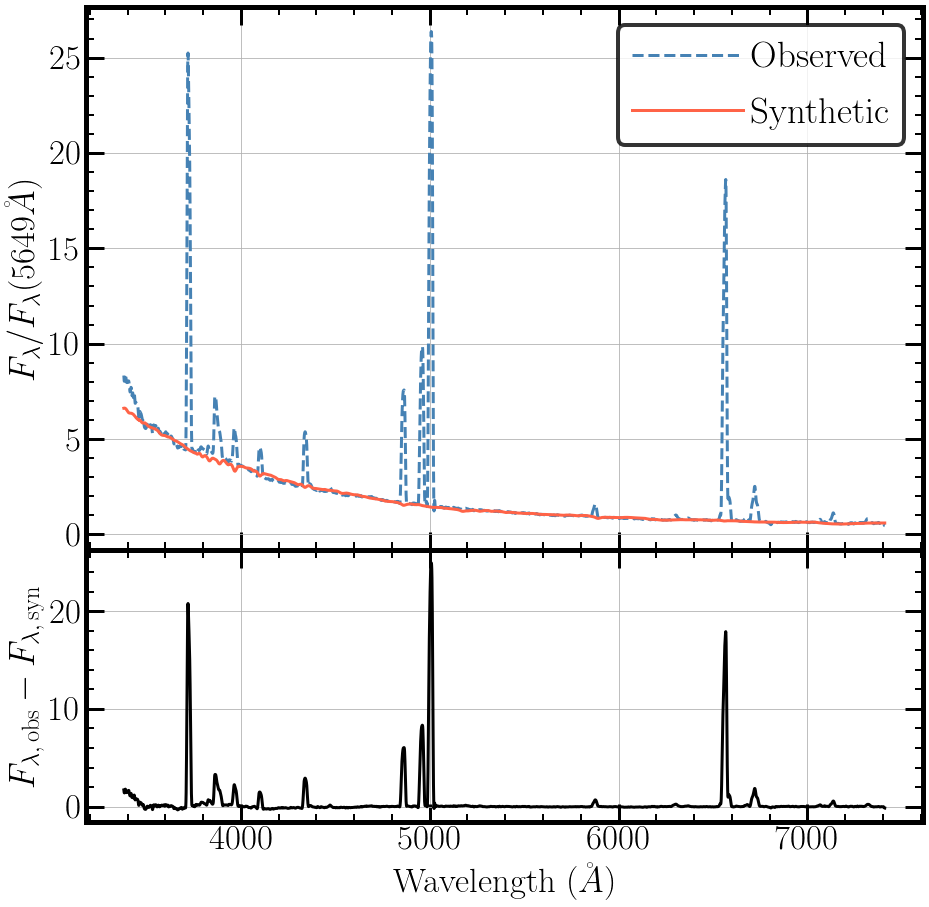}
}
 
\caption{ 
Observed (blue dashed) and model starlight spectra (red solid) for the Region 1 (top row) and Region 2 (bottom row). 
The high  and low-resolution spectra are shown in the left-hand and the right-hand figures respectively. 
The bottom panel of each figure shows the starlight-subtracted  (emission-line) spectrum.
}
     \label{fig:1-11}
\end{center}
\end{figure*}

In general, spectra of galaxies show stellar continuum and strong stellar atmospheric absorption features, including the Balmer lines in addition to any nebular emission component (e.g. Balmer lines that would affect the determination of the stellar component). 
Therefore subtraction of the stellar light is necessary for the correct measurement of the ionized-gas emission-line flux.
We did not attempt to model the nebular continuum component since it is much weaker than the stellar component, and it would not bias the starlight subtraction.
We calculated the stellar component by fitting the spectra for each region with the \texttt{STARLIGHT} code \citep{2005MNRAS.358..363C,2006MNRAS.370..721M}.
We used a base consisting of 150 single stellar populations from the \texttt{BC03} \citep[][]{2003MNRAS.344.1000B} models, with ages ranging between $1$\,Myr--$18$\,Gyr, and metallicity ranging from $\rm Z = 0.0001$--$0.05$.
We also excluded from this analysis the range around strong nebular emission lines. 
The base spectra were convolved with a Gaussian function in order to account for the resolution of the instrument and the velocity dispersion of the galaxy.
In our analysis, because of the low resolution of the spectra obtained with grism \#11, we allowed for a velocity dispersion up to 1200\,km/s.
In Figure~\ref{fig:1-11} we show examples of the resulting stellar light model compared to the observed spectrum, for the high and low resolution spectra of Regions 1 and 2 respectively. 
The bottom panel shows the emission-line spectrum resulting from the subtraction of the stellar component from the observed spectrum. 
These are the spectra used in any subsequent analysis.

\begin{figure*} 
\begin{center}
\hbox{
\includegraphics[width=\columnwidth]{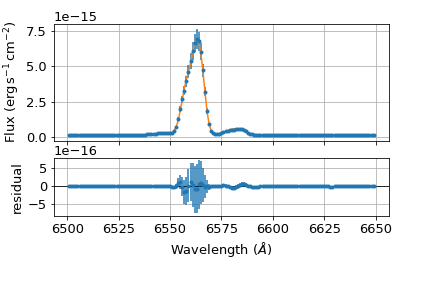}
\hspace{0.7truecm}
\includegraphics[width=\columnwidth]{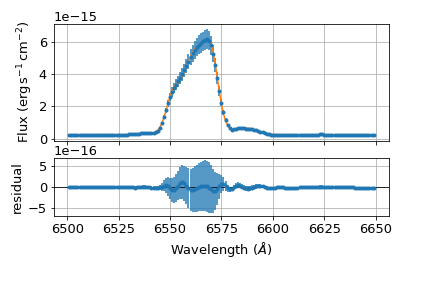}
}
\caption{
The high and low resolution data (blue) and best-fit models (orange) for the region 2, starlight-subtracted spectra, in the wavelength range around the \Ha\ line are presented in the left and right panels respectively. 
The bottom panels present the fit residuals.
}
    \label{fig:Ha_c}
\end{center}
\end{figure*}

We used the \texttt{Sherpa v.4.9} package \citep{2001SPIE.4477...76F, SciPyProceedings_51} in order to fit the emission lines on the starlight-subtracted spectra  and measure their fluxes. 
\texttt{Sherpa} allows us to fit complex models and to determine their parameters and corresponding uncertainties while accounting for measurement uncertainties on the data.
The emission lines of the extracted spectra suffered from non Gaussian shapes in all cases due to instrumentation and setup (e.g. Fig.~\ref{fig:Ha_c}).
In the case of the low resolution spectra, the fact that \Ha\ and the two $[$\ion{N}{II}$]$ emission lines are partially blended added an extra difficulty in measuring the weak [\ion{N}{II}]$\,\lambda 6548$\,\AA~line. 
Although this line is not used in our analysis, accounting for its presence is important for accurately measuring the \Ha\ flux in low-resolution spectra.
In order to account for the complex shape of the lines, we modeled each emission line with three Gaussians. This allowed for more flexibility, which results in better fits in comparison to single Gaussian fits.
We fitted the region of the spectrum around each line of interest separately in order to account for any residual flux variations of the continuum, and variations of the spectral resolution.

\subsection{X-ray data}
\label{sec:xray_spectra}

\cha\ has observed NGC\,922 in two occasions (OBSIDs 10563, 10564; P.I.: A. Prestwich) for a total exposure of 49.7\,ksec. 
The observations were performed with the ACIS-S camera, with the target positioned on the aim-point of the back illuminated ACIS-S3 chip. 
The details of the observations and the analysis of the discrete X-ray sources and ULX populations of NGC\,922 are presented in \cite{2012ApJ...747..150P}.
Because we were interested in the comparison between the X-ray emission, star-forming activity, and metallicity in the regions targeted by the optical spectroscopic observations, we obtained and reanalyzed the \cha\ data for NGC\,922.

X-ray data analysis was performed with \texttt{CIAO}\footnote{\url{http://cxc.harvard.edu/ciao}} \citep{Fruscione06} version 4.12. After the initial processing with the \texttt{chandra\_repro} tool, we extracted images and exposure maps in the broad (0.5--8.0\,keV) band using the \texttt{merge\_obs} tool.
The latter combines data from different observations after re-projecting them to a common reference frame, calculating exposure maps for each observation, and combining the exposure-corrected images. 
To visualize the spatial distribution of the X-ray emission of NGC\,922, the combined image was first adaptively smoothed with the \texttt{csmooth} \texttt{CIAO} tool with a minimum significance of $2\sigma$ and a maximum kernel of 25\,pixels.
Then, the image was exposure-corrected by dividing with the combined broad-band exposure map, which was also smoothed with the same smoothing scales. 
The resulting image is shown in the bottom-right panel of Figure \ref{fig:Maps}.

The X-ray spectra were extracted from the event files using the \texttt{specextract} tool.
Response and ancillary response files were calculated using \texttt{CALDB v.4.9.2.1}.
Our primary regions of interest were those for which we had optical spectra and exhibited X-ray emission:
regions 1 and 3 (referred to as \textit{inner bulge} for the next part of the analysis), and region 2 (referred to as \textit{R2}).
In addition we extracted spectra from larger regions of X-ray emission associated with characteristic features of NGC\,922 (the \textit{bar}, the \textit{bulge}, and the \textit{ring}; Fig. \ref{fig:Slit_positions}). 
The extraction regions (Fig.\,\ref{fig:Maps}) were defined based on the \Ha\ and WISE 12\,$\mu$m maps in order to encompass the star-forming activity in the regions of interest. 
Special care is taken so any X-ray sources associated with each of these regions are fully included.
The background spectrum was measured from source-free regions outside the optical outline of the galaxy. 

The X-ray spectra were binned to include at least 15 counts in each bin, in order to allow the use of $\chi^{2}$ statistics (only in the case of the \textit{inner bulge} and the \textit{bar} which have very few counts we fitted the unbinned spectrum using the \texttt{wstat} statistic). 
For all regions apart from the full galaxy, the background is negligible in the energy range of interest (0.5-8.0\,keV). 
Instead, because of the large extent of the galaxy, the total spectrum includes significant background contamination. 
For this reason the total galaxy spectrum was adaptively binned, so each bin has a S/N of at least 2. 
The spectral analysis was performed with \texttt{Sherpa} in \texttt{CIAO\,v.\,4.12}. 
We used the $\chi^{2}$ statistic for all spectral fits apart from the \textit{inner bulge} and the \textit{bar} regions where, because of the small number of counts, we also fitted the unbinned spectrum using the w-statistic, yielding essentially identical results to the $\chi^{2}$ fit (here we report results from the \texttt{wstat} analysis). 
The reported uncertainties correspond to the 68\% confidence interval for one interesting parameter based on draws of the model parameters from a multivariate normal distribution using the \texttt{confidence} command. 

All spectra were initially fitted with an absorbed power-law model (\texttt{tbabs $\times$ po}) using the \cite{2000ApJ...542..914W} absorption cross-sections. 
This model gave an excellent fit to the spectra of the \textit{bar}, \textit{R2}, and the larger \textit{ring} regions. 
The \textit{bar} and \textit{inner bulge} region gave rather steep photon indices ($\Gamma=3.3_{-1.2}^{+1.9}$ and $\Gamma=4.1_{-1.0}^{+3.1}$ respectively), which indicate a thermal plasma model. Indeed,  an \texttt{APEC} model gave a slightly improved fit for the \textit{inner bulge} spectrum with a best-fit temperature of kT=$0.23^{+0.02}_{-0.08}$\,keV. In the case of the \textit{bar} the thermal plasma model did not improve the quality of the fit. 
The \textit{bulge} and the \textit{total} spectrum of the galaxy showed strong residuals in the 1--2\,keV range indicating an additional thermal component. Indeed, a composite model consisting of a power-law and thermal-plasma component [\texttt{tbabs $\times$ (apec+po)}] gave a significantly better fit.  We also tried this model to the  \textit{bar} and \textit{inner bulge} which show relatively soft spectra.  While the fit of the  \textit{inner bulge} was improved, the fit of the \textit{bar} gave effectively the same fit statistic (wstat).  Therefore, we consider a power-law as the best-fit model for the \textit{bar} spectrum, while for the  \textit{inner bulge}  we adopt the composite thermal plasma - power-law model, in order to obtain a better picture of the power-law luminosity (corresponding to the XRBs component) even under the presence of a weak thermal plasma model.  

We calculated the X-ray flux using the \texttt{sample\_flux} command, which gives the median and the 68\% percentile of the flux distribution based on the model parameter draws from the covariance matrix of the best-fit model. 
In the case of the composite \texttt{apec+po} models we calculated both the total flux and the flux originating only from the power-law component.
The best-fit model parameters and total flux are given in Table\,\ref{tab:X-rays}

\begin{table*}
    \centering
    \caption{X-ray spectral best-fit results for the sub-galactic regions defined in Figure\,\ref{fig:Maps}.}
    \setlength\tabcolsep{2.pt}
    \renewcommand{\arraystretch}{1.4}
	\begin{threeparttable}
    \begin{tabular}{c|c|c|c|c|c|c|c|c|c}
         Region & net counts & $L_X^{\rm 0.5-10 \, keV}$ & ${L_X^{\rm 0.5-8 \, keV}}^{\star}$ & $L_X^{\rm 2-10 \, keV}$ & Model & Red. statistic & $\Gamma$ & N$_H$ & kT \\
         &  & $\rm 10^{40} \, erg \, s^{-1}$ & $\rm 10^{40} \, erg \, s^{-1}$ & $\rm 10^{40} \, erg \, s^{-1}$ & \texttt{Sherpa} & $\chi^{2}$/d.o.f &  & $\rm 10^{22} \, cm^{-2}$ & keV\\
         \hline
         \textit{total} & 1685.8 & $6.49^{+1.68}_{-1.92}$ & $5.92^{+1.48}_{-1.75}$ & $3.54^{+0.97}_{- 1.12}$ & PO+APEC & 65.8/112 & $1.99^{+0.25}_{-0.21}$ & $0.22^{+0.22}_{-0.12}$ & $0.23^{+0.04}_{-0.04}$\\
         \textit{bulge} & 306.3 & $1.11^{+0.62}_{-0.52}$ & $1.02^{+0.56}_{-0.47}$ & $0.50^{+0.27}_{-0.25}$ & PO+APEC & 5.81/16 & $1.90^{+0.41}_{-0.36}$ & $0.31^{+0.22}_{-0.19}$ & $0.23^{+0.06}_{-0.05}$\\
         \textit{inner bulge} & 81.5 & $0.36^{+0.65}_{-0.35}$ & $0.27^{+0.35}_{-0.24}$ & $0.11^{+0.64}_{-0.03}$ & PO+APEC & 164.9/514$^{\dag}$ & $-0.34^{+1.03}_{-0.82}$ & $0.51^{+0.10}_{-0.11}$ & $0.19^{+0.03}_{-0.02}$\\
        \textit{bar} & 75.0 & $0.24^{+0.03}_{-0.05}$ & $0.23^{+0.03}_{-0.04}$ & $0.09^{+0.03}_{-0.03}$ & PO & 214.1/511$^{\dag}$ & $2.43^{+0.37}_{-0.24}$ & ${0.16^{+0.06}_{- \ldots}}^{\P}$  & -  \\
         \textit{R2} & 456.6 & $2.12^{+0.44}_{-0.55}$  & $1.89^{+0.09}_{-0.14}$ & $1.45^{+1.11}_{-0.12}$ & PO & 16.5/27 & $1.74^{+0.21}_{-0.19}$ & $0.10^{+0.06}_{-0.05}$ & - \\
         \textit{ring} & 819.8 & $3.55^{+0.63}_{-0.78}$ & $3.16^{+0.50}_{-0.64}$ & $2.32^{+0.52}_{-0.63}$ & PO & 38.6/55 & $1.76^{+0.18}_{-0.17}$ & $0.05^{+0.05}_{-0.05}$ & - \\
    \end{tabular}
    \label{tab:X-rays}
    \begin{tablenotes}
        \item $^{\star}$ The $0.5$--$8$\,keV luminosities refer to the power-law component only.
        \item $\dag$ The fit was performed on the unbinned data using the \texttt{wstat} statistic. 
        \item $\P$ The parameter pegged at the low bound (the Galactic line-of-sight \ion{H}{I} column density).
    \end{tablenotes}
    \end{threeparttable}
\end{table*}

\section{Results}
\label{sec:Results}

\subsection{SFR and stellar mass maps}

In order to examine the metallicity-SFR-$L_X$ relation at the sub-galactic level, we created spatially resolved SFR, stellar mass, and X-ray emission maps.

 As a stellar mass indicator, we used the 3.4\,$\mu$m WISE band-1 data retrieved from the Infrared Science Archive (IRSA\footnote{\url{https://irsa.ipac.caltech.edu/frontpage/}}).
We converted the 3.4\,$\mu$m WISE band-1 luminosity to stellar mass using the conversion of  \cite{2013MNRAS.433.2946W}: 

\begin{eqnarray}
    \rm log \frac{\textit{M}_\star}{M_\odot}=-0.040 + 1.12 \, log \frac{\nu \textit{L}_{\nu}(3.4\,\mu m)}{\textit{L}_\odot}\quad.
    \label{eq:Mstar}
\end{eqnarray}

As a SFR indicator, we used MIPS 24\,$\mu$m \textit{Spitzer} data which probe dust heated by young stellar populations. We retrieved the post-BCD (post-Basic Calibrated Data) from the IRSA archive.
The MIPS 24\,$\mu$m luminosity was converted to SFR using the
calibration of \cite{2007ApJ...666..870C}

\begin{eqnarray}
   \frac{\rm SFR_{24\micron}}{(\msunpyr)} = 1.27 \times 10^{-38}
    \bigg[\frac{{L}_{24\micron}}{(\erg)} \bigg]^{0.8850} \quad.
	\label{eq:SFR-24}
\end{eqnarray}

As an alternative SFR indicator we used the 12\,$\mu$m WISE W3 band data (also obtained from the IRSA archive), which are dominated by emission by Polycyclic Aromatic Hydrocarbons (PAH). 
The  WISE  band-3 luminosity was  converted to SFR using  the \cite{2017ApJ...850...68C} calibration.

\begin{eqnarray}
    \rm log \, \frac{\rm SFR_{12\micron}}{(\msunpyr)} = 0.889 \, log \, \frac{\textit{L}_{12\mu m} }{\textit{L}_\odot} - 7.76\quad.
	\label{eq:SFR-W3}
\end{eqnarray}

\begin{figure*}
\begin{center}
\includegraphics[width=\linewidth]{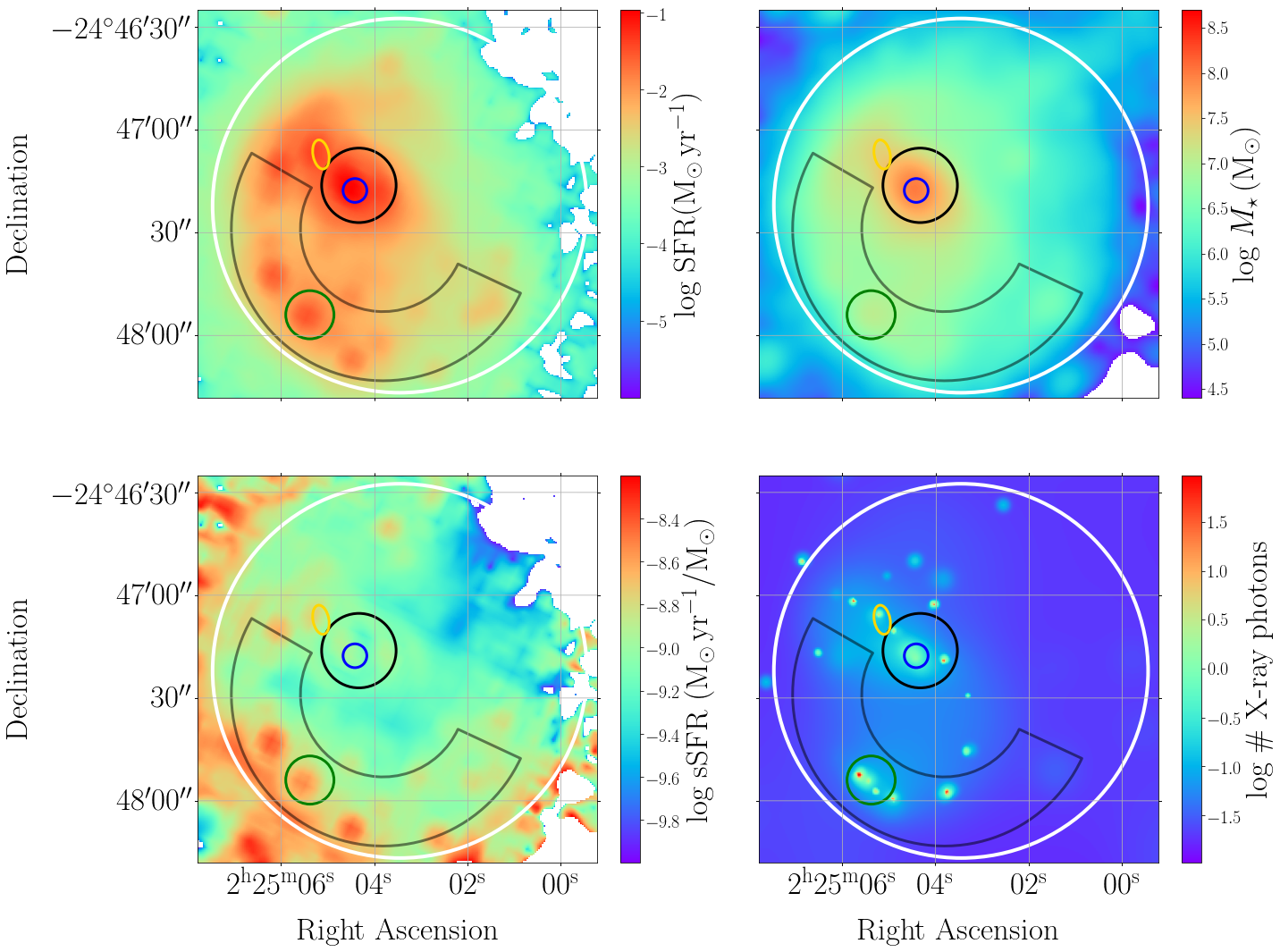}
\caption{NGC\,922 maps of SFR (top left) as measured from MIPS 24$\mu$m observations, stellar mass from WISE band\,1 (top right), sSFR (bottom left), and X-ray intensity in the 0.5--8.0\,keV band (bottom right). The \textit{total} galaxy is defined by the white circle. The other regions define the \textit{bulge} (black circle), \textit{inner bulge} (blue circle), \textit{bar} (yellow ellipse), and \textit{R2} (green circle). The gray semi-annulus shows the \textit{ring} region.}
    \label{fig:Maps}
\end{center}
\end{figure*}

The stellar mass, SFR, specific SFR (SFR/$M_\star$; sSFR) and X-ray emission maps (based on the exposure map corrected full-band images described in Section \ref{sec:xray_spectra}) of NGC\,922 are presented in Figure\,\ref{fig:Maps}, as well as the regions we defined and compared in the following.

\subsection{Optical spectra}
\label{sec:Optical_results}

The emission-line fluxes for the high and low-resolution spectra from each region of NGC\,922 (as defined in Fig.\,\ref{fig:Slit_positions} and Table \ref{tab:Obs}) are presented in Table\,\ref{tab:lines}. 
As we see from the table, there is a systematic difference between the high and low resolution spectral extractions, where the former show slightly higher fluxes (0.1-0.3~dex) in the blue region ($\sim4861$\,\AA) of the spectra. 
Because our project required observations of NGC\,922 with particular slit rotations (Section \ref{sec:Observations}), the slit positions angle was not aligned with the parallactic angle. 
This difference can cause loss of flux in the blue part of the spectrum because of  differential diffraction on the observed spectra, especially for observations at high airmass \citep[e.g.][]{1982PASP...94..715F}.
The problem cannot be fully remedied by the flux-calibration process since the standard stars were observed at not negligible average airmass (${\simeq} 1.4$) leading to flux loss in the blue area of the spectrum, and subsequently higher flux in the blue part of the flux-calibrated object spectra. 
This effect also resulted in slightly different absolute flux calibrations between the high and low resolution spectra. 
Despite this systematic difference, the  emission-line fluxes measured for the same regions are consistent within the measurement uncertainties.
Because of the parallactic angle effects, we could not use the Balmer decrement to reliably measure the extinction in the different regions.
However, this problem did not affect the accuracy of the emission-line ratios presented here, because these ratios involve emission lines that are nearby in wavelength. 

\begin{table*}
	%\scriptsize
	%\footnotesize
	\centering
	\caption{Logarithm of the emission-line fluxes (in units of $\rm erg \, s^{-1} \, cm^{-2}$).}
	\label{tab:lines}
	\setlength\tabcolsep{5.pt}
	\begin{threeparttable}
	\begin{tabular}{c|c|c|c|c|c|c|c|c|c|c|c|c|c}
	    Ion & Line & \multicolumn{12}{c}{Region} \\
      & & 1 & 1 & 2 & 2 & 3 & 3 & 4 & 4 & 5 & 6 & 7 & 8\\
     & ($\lambda$ \AA) & low & high & low & high & low & high & low & high & low & low & low & low\\ 
    \hline
    $[$\ion{O}{II}$]$ & $3726/9$ & $ -13.8$ & $\ldots$ & $ -13.8$ & $\ldots$ & $ -13.5$ & $\ldots$ & $ -14.6$ & $\ldots$ & $ -14.2$ & $ -14.2$ & $ -14.8$ & $ -14.1$\\
    H$\gamma$ & $4340$ & $ -15.1$ & $\ldots$ & $ -14.6$ & $\ldots$ & $ -14.6$ & $\ldots$ & $ -15.6$ & $\ldots$ & $ -15.1$ & $ -15.3$ & $ -15.8$ & $ -14.9$\\
    \Hb & $4861$ & $ -14.4$ & $ -14.2$ & $ -14.4$ & $ -14.2$ & $ -14.2$ & $ -13.9$ & $ -15.2$ & $ -15.0$ & $ -14.8$ & $ -14.8$ & $ -15.5$ & $ -14.6$\\
   $[$\ion{O}{III}$]$ & $4959$ & $ -15.0$ & $ -14.9$ & $ -14.3$ & $ -14.0$ & $ -14.7$ & $ -14.6$ & $ -15.3$ & $ -15.1$ & $ -14.9$ & $ -15.0$ & $ -15.9$ & $ -14.8$\\
   $[$\ion{O}{III}$]$ & $5007$ & $ -14.6$ & $ -14.6$ & $ -13.7$ & $ -13.5$ & $ -14.2$ & $ -14.2$ & $ -14.8$ & $ -14.7$ & $ -14.4$ & $ -14.5$ & $ -15.5$ & $ -14.2$\\
    \ion{He}{I} & $5876$ & $ -15.3$ & $ -15.4$ & $ -15.4$ & $ -15.4$ & $ -15.0$ & $ -15.0$ & $ -16.5^{*}$ & $ -16.1^{*}$ & $ -15.8$ & $ -15.8^{*}$ & $ -16.6^{*}$ & $ -15.5$\\
    $[$\ion{O}{I}$]$ & $6046$ & $ -15.5$ & $ -15.8$ & $ -15.9$ & $ -15.8$ & $ -15.4$ & $ -15.3$ & $ -16.4^{*}$ & $ -16.3^{*}$ & $ -16.0^{*}$ & $ -15.7^{*}$ & $ -16.5^{*}$ & $ -15.6$\\
    $[$\ion{O}{I}$]$ & $6300$ & $ -16.1^{*}$ & $ -16.4^{*}$ & $ -16.3^{*}$ & $ -16.3^{*}$ & $ -15.6^{*}$ & $ -15.7^{*}$ & $ -16.9^{*}$ & $ -16.5^{*}$ & $ -16.4^{*}$ & $ -16.3^{*}$ & $ -17.3^{*}$ & $ -16.1^{*}$\\
    $[$\ion{N}{II}$]$ & $6548$ & $ -15.4^{\dag}$ & $ -15.2$ & $ -15.7^{\dag}$ & $ -15.6$ & $ -15.0^{\dag}$ & $ -14.9$ & $ -16.0^{\dag}$ & $ -16.3$ & $ -16.3^{\dag}$ & $ -16.1^{\dag}$ & $ -17.3^{\dag}$ & $ -15.3^{\dag}$\\
    \Ha\ & $6563$ & $ -14.0$ & $ -14.0$ & $ -13.9$ & $ -14.0$ & $ -13.6$ & $ -13.6$ & $ -14.8$ & $ -14.8$ & $ -14.3$ & $ -14.5$ & $ -15.1$ & $ -14.2$\\
    $[$\ion{N}{II}$]$ & $6583$ & $ -14.7$ & $ -14.6$ & $ -15.0$ & $ -15.2$ & $ -14.4$ & $ -14.3$ & $ -15.8$ & $ -15.7$ & $ -15.5$ & $ -15.5$ & $ -15.9$ & $ -15.2$\\
    $[$\ion{S}{II}$]$ & $6716/31$ & $ -15.2$ & $\ldots$ & $ -15.5$ & $\ldots$ & $ -14.9$ & $\ldots$ & $ -16.3$ & $\ldots$ & $ -15.7$ & $ -15.7$ & $ -16.1$ & $ -15.6$\\
    \ion{He}{I} & $7065$ & $ -16.0^{*}$ & $\ldots$ & $ -16.2^{*}$ & $\ldots$ & $ -15.4^{*}$ & $\ldots$ & $ -16.7^{*}$ & $\ldots$ & $ -16.3^{*}$ & $ -16.3^{*}$ & $ -16.8^{*}$ & $ -16.3^{*}$\\
    $[$\ion{Ar}{III}$]$ & $7136$ & $\ldots$ & $\ldots$ & $ -15.5$ & $\ldots$ & $ -15.3^{*}$ & $\ldots$ & $ -16.3^{*}$ & $\ldots$ & $ -15.9^{*}$ & $ -16.0^{*}$ & $ -16.6^{*}$ & $ -15.9^{*}$\\
    $[$\ion{O}{II}$]$ & $7320/31$ & $ -17.0^{*}$ & $\ldots$ & $ -16.1$ & $\ldots$ & $ -15.4^{*}$ & $\ldots$ & $ -17.2^{*}$ & $\ldots$ & $ -16.2^{*}$ & $ -16.0^{*}$ & $ -16.0^{*}$ & $ -16.6^{*}$\\
	\end{tabular}
	\begin{tablenotes}
	    \item The emission line fluxes show uncertainty log\,$f \rm \simeq \pm0.1 \, erg \, s^{-1} \, cm^{-2}$ unless otherwise indicated.
	    \item $^{*}$ Noise dominates the spectrum in the region of the particular emission line, therefore the flux estimation is highly uncertain.
	    \item $^{\dag}$ Large uncertainty due to partial blending with \Ha\ caused by low resolution.
    \end{tablenotes}
	\end{threeparttable}
%	\end{small}
\end{table*}

One thing that is clear from the analysis of the spectra for the different star-forming region, is that they all show prominent Balmer and \ion{He}{I}\,($\lambda \, 5876$\AA) lines regardless of their location in the galaxy (bulge, ring, or intermediate region). The presence of the \ion{He}{I} line in particular indicates ionization by strong UV continuum that can be produced by very young stellar populations, or, in the case of the bulge regions, by a potential active galactic nucleus (AGN). 

\begin{figure*}
\begin{center}
\includegraphics[width=\linewidth]{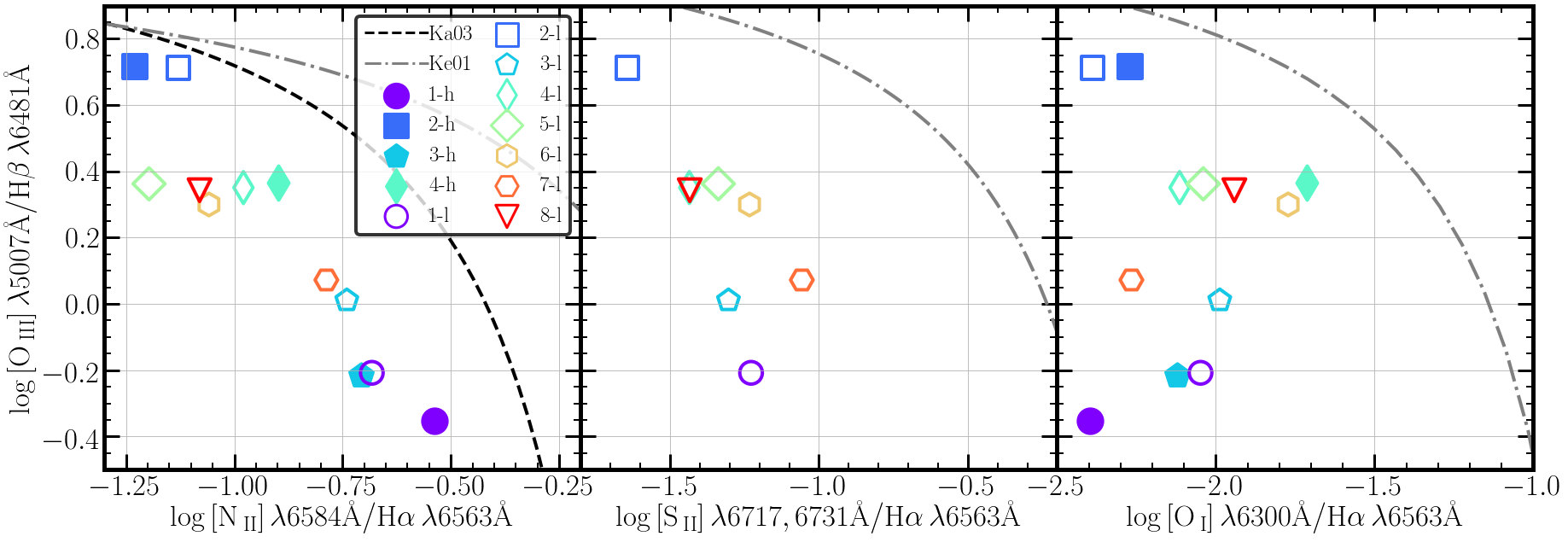}
\caption{From left to right are the ($\rm[\ion{N}{II}]$/\Ha--$[$\ion{O}{III}$]$/\Hb), ($\rm[\ion{S}{II}]$/\Ha--$[$\ion{O}{III}$]$/\Hb), and ($\rm[\ion{O}{I}]$/\Ha--$[$\ion{O}{III}$]$/\Hb) BPT diagrams for all the regions (IDs in legend) and spectrum resolutions (\textit{h} and \textit{l} refers to high and low resolution respectively). 
The \protect\cite{2001ApJ...556..121K} and  \protect\cite{2003MNRAS.346.1055K} curves separating regions excited by non-stellar photoinization (top right) from \ion{H}{II} regions (lower left) are presented with gray dashed-doted and black dashed lines respectively. 
All regions are well within the locus of \ion{H}{II} regions.}
    \label{fig:BPT}
\end{center}
\end{figure*}

The location of the sources on line-ratio diagnostic diagrams \citep[BPT diagrams;][]{1981PASP...93....5B,1987ApJS...63..295V} allows us to infer the source of their excitation: photoionization by stellar populations (star-forming regions), photoionization by non-stellar continuum (AGN), or shock excitation (e.g. from supernovae or strong stellar winds from massive stars). 
In Figure\,\ref{fig:BPT} we present the location of the different regions (Fig. \ref{fig:Slit_positions}) on the ($\rm[\ion{N}{II}]$/\Ha--$[$\ion{O}{III}$]$/\Hb), ($\rm[\ion{S}{II}]$/\Ha--$[$\ion{O}{III}$]$/\Hb), and ($\rm[\ion{O}{I}]$/\Ha--$[$\ion{O}{III}$]$/\Hb) BPT diagrams.
All the NGC\,922 sub-galactic regions examined here are encompassed by the theoretical \cite{2001ApJ...556..121K} and the empirical \cite{2003MNRAS.346.1055K} curves which delineate star-forming region from AGN and shock-dominated regions. 
We see that none of the regions in the central part of the bulge lies in the AGN locus, indicating that NGC\,922 does not host an AGN. 
In addition none of the regions in the bulge, the ring, or the intermediate regions shows evidence for strong shock excitation indicating that the dominant source of ionization are the young stellar populations.
Regions located in the ring and the bulge show distinct $\rm[\ion{N}{II}]$/\Ha\ and $[$\ion{O}{III}$]$/\Hb\ emission line ratios and reside in clearly separate loci of the $\rm[\ion{N}{II}]$ BPT diagram. 
Regions located on the ring of NGC\,922 are in the upper left of the diagram, while regions located on the bulge, or intermediate locations (like region 7), reside on the lower right part of the BPT diagram.
$[$\ion{S}{II}$]$/\Ha, and $[$\ion{O}{I}$]$/\Ha\ emission line ratios do not distinguish the bulge and ring regions as well as the $\rm[\ion{N}{II}]$/\Ha\ and $[$\ion{O}{III}$]$/\Hb\ ratios.

\subsection{Metallicity measurements and stellar population parameters}

Metallicities were derived from the fluxes reported in Table \ref{tab:lines} using calibrations provided by \cite{2004MNRAS.348L..59P}:
\begin{eqnarray}
    \rm [12 + log(O/H)] =  8.90 + 0.57 \times N2
\end{eqnarray}
\begin{eqnarray}
    \rm [12 + log(O/H)] =  8.73 - 0.32 \times O3N2
\end{eqnarray}
where
\begin{eqnarray}
    \rm  N2 = log(\textit{f}_{[\ion{N}{II}]_{\lambda \, 6583}}/\textit{f}_{H\alpha_{\lambda \, 6563}})
\end{eqnarray}
\begin{eqnarray}
   \rm O3N2 = log \, \frac{\textit{f}_{[O\,_{III}]_{\lambda \, 5007}}/\textit{f}_{H\beta_{\lambda \, 4863}}} {\textit{f}_{[\ion{N}{II}]_{\lambda \, 6583}}/\textit{f}_{H\alpha_{\lambda \, 6563}}}
\end{eqnarray}

 where \textit{f} corresponds to each emission-line flux. 
 These diagnostics are well calibrated for the range of metallicities we find \citep{2008ApJ...681.1183K}.

\begin{figure*}
    \begin{center}
    \includegraphics[width=\linewidth]{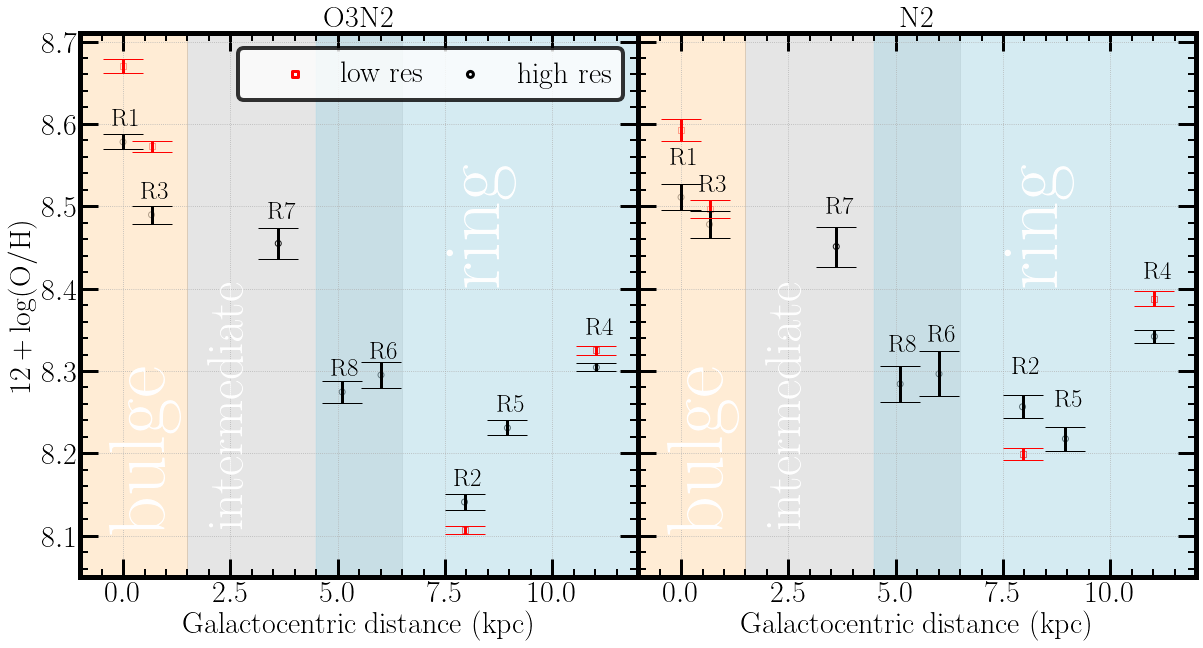}
    \caption{Gas phase metallicities for all the regions as a function of their galactocentric distance (adopting as center of the galaxy the center of region 1).
    The corresponding region IDs (Fig.\,\ref{fig:Slit_positions}) are shown on top of the points. 
    High resolution extractions are presented with red colour and low with black. 
    In the left and right panels are the O3N2 and N2 diagnostic results respectively.
    The orange, gray, and blue shaded areas represent regions encompassed in the bulge, intermediate, and ring loci of NGC\,922 respectively.
    Due to the asymmetrical shape of NGC\,922, the shading overlaps in a range of radii. This is clearly seen by comparing the position of region 8, which is located in the north-west part of the ring, that is relatively closer to bulge, with that of region 6 which does not belong to the ring.}
    \label{fig:Metal}
    \end{center}
\end{figure*}

Metallicity measurements for all spectral extractions using both methods are presented in Table\,\ref{tab:Metal} and Figure\,\ref{fig:Metal}.
The metallicity of different regions within NGC\,922 ranges from near-solar to sub-solar across the galaxy.
We find that regions located on the ring (2, 4, 5, and 8) show consistently significantly lower metallicity ($\rm 8.11<[12+log(O/H)]<8.39$) compared to regions located on the bulge (regions 1 and 3; $\rm 8.49<[12+log(O/H)]<8.67$) regardless of the measurement method (the quoted ranges include the full range of both methods).
Region 7, which is located between the ring and the bulge, shows an intermediate value ($\rm [12+log(O/H)]=8.45$).
The metallicity of region 6, that is not placed on the ring, is $\rm [12+log(O/H)]=8.30$.

Metallicities derived with the O3N2 diagnostic show wider differences between the ring and the bulge compared to the N2 diagnostic (Fig. \ref{fig:Metal}).
Similarly, the high-resolution extractions tend to extend the differences in both diagnostics. This is because of the better determination of the Balmer-line fluxes resulting from the more accurate subtraction of the stellar component and modelling of the \Ha-[\ion{N}{II}] complex. 

The metallicities measured for the regions located in the bulge (regions 1 and 3; $\rm [12+log(O/H)]=8.50$--$8.67$) are in agreement with the one reported in \cite{2006MNRAS.370.1607W} for a 6.7" diameter region on the bulge, based on the N2 method ($\rm [12+log(O/H)]=8.6$).
The same work also reports metallicities based on the $[$\ion{N}{II}$]$/$[$\ion{S}{II}$]$ ratio ($\rm [12+log(O/H)] \simeq 9$) and the R-band luminosity-metallicity relation of \citealt{2004MNRAS.350..396L}, which, however, are  higher than those found from the N2O3 or the N2 methods.  
This over-estimation could be due to the fact that the $[$\ion{N}{II}$]$/$[$\ion{S}{II}$]$ ratio becomes insensitive to metallicity at low metallicities and the $[$\ion{N}{II}$]$/$[$\ion{S}{II}$]$-metallicity relation is a sensitive function of the ionization parameter \citep[e.g.][]{2013ApJS..208...10D,Blanc_2014} which is not known for the different star-forming regions in NGC\,922.
Our measurements of the bulge metallicity are also in agreement with the one reported in \citealt{2013ApJ...773....4R} ($\rm [12+log(O/H)] = 8.75 \pm 0.08$).

\begin{table}
	\centering
	\caption{Metallicity measurements}
	\label{tab:Metal}
	\begin{tabular}{c|c|c|c}
    \hline\hline
    Region &  Grism & \multicolumn{2}{c}{Metallicity} \\
	ID & \# & N2 & O3N2\\
	& & 12 + log(O/H) & 12 + log(O/H)\\
    \hline
    1  &  11  & $ 8.51 {\pm} 0.02 $ & $ 8.58 {\pm} 0.01 $ \\
    1  &  18  & $ 8.59 {\pm} 0.01 $ & $ 8.67 {\pm} 0.01 $ \\
    2  &  11  & $ 8.26 {\pm} 0.01 $ & $ 8.14 {\pm} 0.01 $ \\
    2  &  18  & $ 8.20 {\pm} 0.01 $ & $ 8.11 {\pm} 0.01 $ \\
    3  &  11  & $ 8.48 {\pm} 0.02 $ & $ 8.49 {\pm} 0.01 $ \\
    3  &  18  & $ 8.50 {\pm} 0.01 $ & $ 8.57 {\pm} 0.01 $ \\
    4  &  11  & $ 8.34 {\pm} 0.01 $ & $ 8.30 {\pm} 0.01 $ \\
    4  &  18  & $ 8.39 {\pm} 0.01 $ & $ 8.33 {\pm} 0.01 $ \\
    5  &  11  & $ 8.22 {\pm} 0.01 $ & $ 8.23 {\pm} 0.01 $ \\
    6  &  11  & $ 8.30 {\pm} 0.03 $ & $ 8.30 {\pm} 0.02 $ \\
    7  &  11  & $ 8.45 {\pm} 0.02 $ & $ 8.45 {\pm} 0.02 $ \\
    8  &  11  & $ 8.28 {\pm} 0.02 $ & $ 8.27 {\pm} 0.01 $ \\
	\end{tabular}
\end{table}

In Table\,\ref{tab:models} we present the SFR, stellar mass, sSFR, and metallicity measurements for five sub-galactic regions and the 
whole NGC\,922 as defined in Figure\,\ref{fig:Maps}. 
Here we adopt the O3N2 metallicity calibration as it is considered more robust especially in star-forming galaxies \citep{2004MNRAS.348L..59P}. 
As a metallicity for the \textit{total} galaxy we used the median value of all our metallicity  measurements (Table\,\ref{tab:Metal}), and as uncertainty the standard deviation of this distribution. 
The extracted regions are evenly spread over the body of the galaxy thus the metallicity of the galaxy is not biased towards a particular region.
Due to the lack of metallicity measurements in the \textit{bar} we adopted the same value and uncertainty as the \textit{total}. For the other sub-galactic regions we adopted the metallicity of the extraction regions they encompass: 
median of regions 1 and 3 for \textit{bulge} and \textit{inner bulge}, region 2 for region \textit{R2}, and the median metallicity (and standard deviation) of extraction regions 2, 4, 5, and 8 for the \textit{ring}.

Our estimation of the total stellar mass of NGC\,922 ($\rm 2.78 \, \pm 0.01 \times 10^{10} \, M_\odot$) is in agreement with the one reported in \citealt{2010AJ....139.1369P} ($\rm 2.8 \times 10^{10} \, M_\odot$).
Our estimation of the integrated SFR of NGC\,922 ($\rm 8.6 \, \pm 0.3 \, M_\odot \, yr^{-1}$) is in agreement with the one reported in \citealt{2018MNRAS.476.5681E} ($\rm 8.5 \, \pm 0.6 \, M_\odot \, yr^{-1}$) which was also calculated through 24\,$\mu$m emission.
This measurement is slightly higher but consistent within the uncertainties with that reported in \citealt{2006MNRAS.370.1607W} (SFR$_{\rm H\alpha}=8.20 \pm 0.32$, SFR$_{\rm UV}=7.04 \pm 0.02 \, {\rm M_\odot \, yr^{-1}}$). 
WISE W3 12\,$\mu$m emission results in a slightly lower total SFR$_{\rm 12 \mu m}$ ($\rm 6.11 \pm 0.04 \, M_\odot \, yr^{-1}$).
While the \Ha-based SFR indicator is a better probe of the stellar populations associated with high-mass XRBs \citep[HMXBs;][]{2020MNRAS.494.5967K}, and therefore correlates better with their X-ray emission, its use at sub-galactic scales is subject to differential extinction within the galaxy. 
In order to correct for the varying extinction between star-forming regions we would need extinction maps for NGC\,922 which are not available. 
Therefore, in our sub-galactic region analysis, we adopted the IR SFR indicators to avoid biases and scatter due to the lack of spatially resolved extinction measurements.

\begin{figure}
    \centering
    \includegraphics[width=\columnwidth]{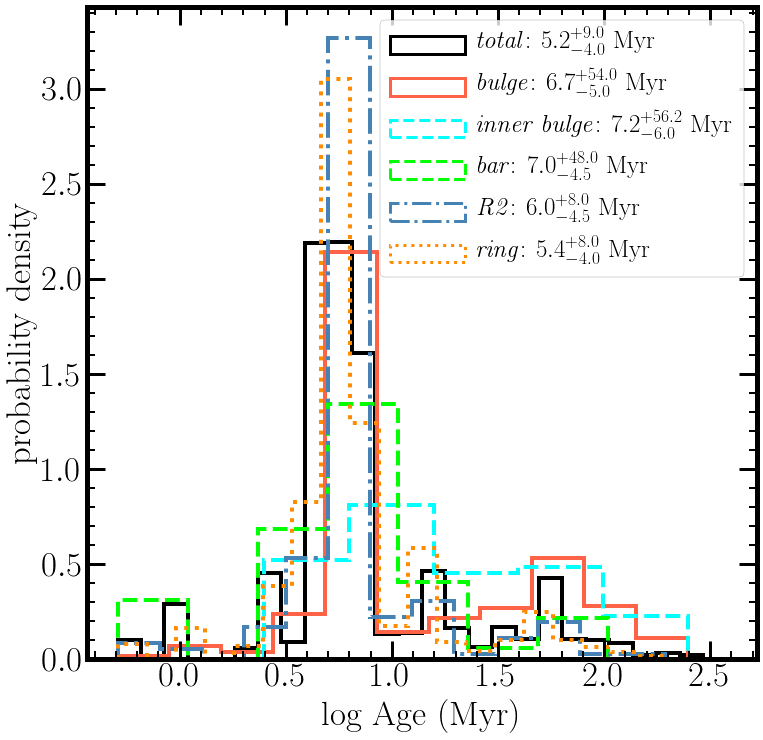}
    \caption{
    Distribution of the star clusters ages from the work of \protect\cite{2010AJ....139.1369P} for the regions defined in Figure \ref{fig:Maps}.
    The modes of the age distributions and the 68\% confidence intervals  for each sub-galactic region are shown in the legend.
    }
    \label{fig:SC_ages_dist}
\end{figure}

\cite{2010AJ....139.1369P} reported the ages of star-clusters in different regions throughout the galaxy. Based on the ages of the individual clusters encompassed within each sub-galactic region, we find the representative age of the stellar populations in these regions.  Figure \ref{fig:SC_ages_dist} shows the distribution of the ages in the star-clusters within each sub-galactic region. We clearly see that all distributions peak at ages below $\sim10$\,Myr (although the \textit{bulge} and the \textit{total} galaxy show a secondary weaker peak at ages $\sim50$\,Myr). 
The ages of the dominant cluster populations in each region are given in Table \ref{tab:models}, along with their corresponding 68\% percentiles.

\subsection{X-ray data results}
\label{sec:X_ray_results}

\begin{figure*}
\begin{center}
\includegraphics[width=0.7\linewidth,angle=270]{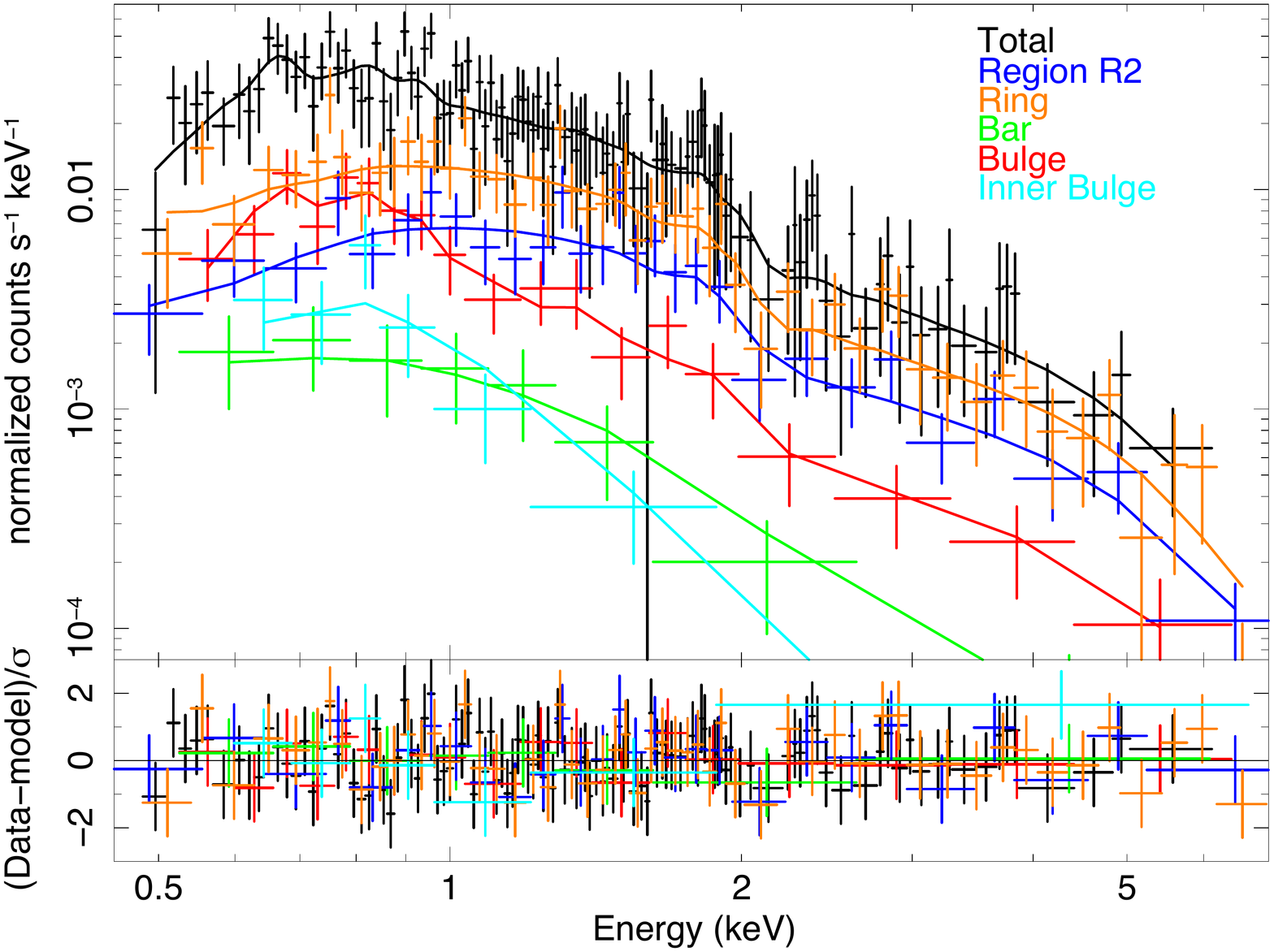}
\caption{X-ray spectra and best-model fits for the \textit{total} emission, \textit{R2}, \textit{ring}, \textit{bar}, \textit{bulge}, and \textit{inner bulge} of NGC\,922 as defined in Fig.\,\ref{fig:Maps} with black, blue, orange, green, red, and light blue colours respectively. The best-model fits for each spectrum are over-plotted with the same colours. Details of the best-model fits are given in Table\,\ref{tab:X-rays}. The fit for the {\textit{inner bulge}}  and the {\textit{bar}} were performed on the unbinned spectrum, here for clarity of presentation the data have been binned to have at least 10 counts be bin. 
In the lower panel shows the fit residuals is units of the data uncertainties ($\sigma$). }
    \label{fig:X-ray_spectra}
\end{center}
\end{figure*}

The X-ray spectra for the different regions (Fig. \ref{fig:Maps}) are very well fitted $\chi^2_{\nu} \leq 0.7$ with either simple power-law or composite power-law thermal-plasma models. The best-fit model parameters and total flux are given in Table\,\ref{tab:X-rays} while the X-ray spectra for each region and the corresponding best-fit models are shown in Figure\,\ref{fig:X-ray_spectra}.
All regions, except from the bulge, have \ion{H}{I} column density slightly higher (but consistent) with the Galactic \nh\ along the line of sight to NGC\,922 \citep[\nh\,$=1.6\times10^{20}$\cmsq;][]{2016AA...594A.116H} based on the \nh\ tool\footnote{\url{https://heasarc.gsfc.nasa.gov/cgi-bin/Tools/w3nh/w3nh.pl}}.
In the full band the thermal component contributes less than 20\% of the total emission of the composite \texttt{po+apec} spectral fits, apart from the \textit{inner bulge} where it contributes $\sim45$\%  of the total 0.5--10\,keV emission.
However, for all regions the thermal component has negligible contribution in the 2--10\,keV X-ray luminosities that are used in the following discussion.
Finally, we note that we do not find any point-like source above our detection limit of $\sim10^{39}$\erg\ that could indicate the presence of an AGN. 
This is in agreement with the non-detection of an AGN-like source in the optical spectra of the central part of the bulge.

\begin{figure}
    \centering
    \includegraphics[width=\columnwidth]{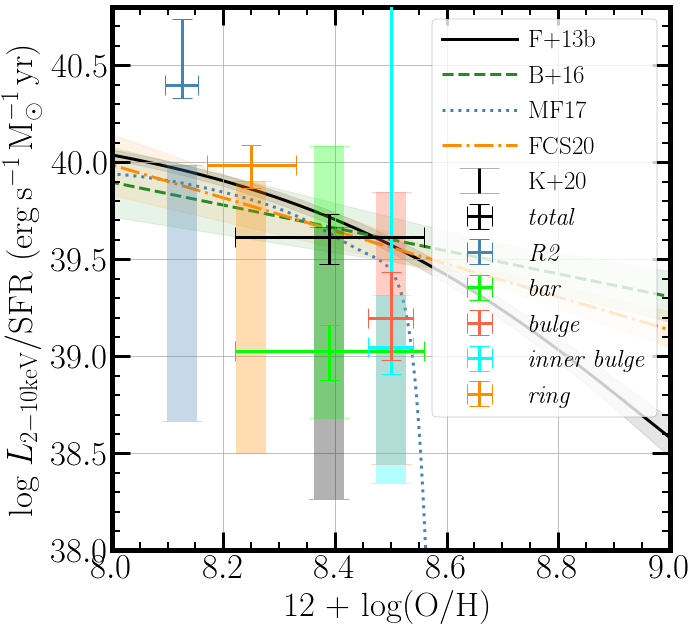}
    \caption{
    X-ray luminosity (2--10\,keV) normalized by SFR as a function of metallicity for five sub-galactic regions and the integrated emission of NGC\,922, as defined in Fig.\,\ref{fig:Maps}.
    Black continuous and blue dotted lines represent the theoretical relations of  \protect\cite{2016ApJ...827L..21F} and \protect\cite{2017ApJ...840...39M} respectively. The green dashed and orange dash-dot lines show the observational relations of \protect\cite{2016MNRAS.457.4081B} and   \protect\cite{2020MNRAS.495..771F}  respectively, while the shaded areas indicate the uncertainties of the corresponding relations.
    %The \protect\cite{2017ApJ...840...39M} model has been converted from \protect\cite{2004ApJ...617..240K} R$_{23}$ to \protect\cite{2004MNRAS.348L..59P} O3N2 using the calibration of \protect\cite{2008ApJ...681.1183K}.
    Stripes with the same colour with the datapoints represent the expected $L_X$/SFR (in the 2--8\,keV band) ratio based on the sub-galactic $L_X$--SFR relation of \protect\cite{2020MNRAS.494.5967K} for the 24\,$\mu$m SFR indicator. 
    This is calculated for each individual sub-galactic region given their SFR and size.
    The wide range of the stripes indicates the 1\,$\sigma$ uncertainty of the scaling relation, including the term which describes the intrinsic scatter of the sub-galactic relation.}
    \label{fig:LX_SFR}
\end{figure}

In Figure\,\ref{fig:LX_SFR} we compare the $L_X$/SFR as a function of metallicity for the considered sub-galactic regions and the total NGC\,922 emission. 
For this comparison we adopted the MIPS 24\,$\mu$m-based SFRs. 
The \textit{total} galaxy X-ray emission is in agreement with the theoretical models of \cite{2016ApJ...827L..21F}, and \cite{2017ApJ...840...39M}, and the empirical fits of \cite{2016MNRAS.457.4081B}, and \cite{2020MNRAS.495..771F}.
For this comparison we converted the \cite{2017ApJ...840...39M} model from the R$_{23}$ metallicity \citep[][]{2004ApJ...617..240K} to O3N2 \cite[][]{2004MNRAS.348L..59P} using the calibration of \cite{2008ApJ...681.1183K}.
We find that the \textit{bulge}, \textit{inner bulge}, and \textit{bar} regions  follow the \cite{2017ApJ...840...39M} model considering the uncertainties.
All regions except \textit{R2} and the \textit{ring} are in the expected range by the sub-galactic $L_X{\sim}{ \rm SFR_{24\,\mu m}}$ fit of \cite{2020MNRAS.494.5967K} considering their 1 $\sigma$ uncertainty and  scatter term. 
\cite{2020MNRAS.494.5967K} have calculated $L_X{\sim}{ \rm SFR}$ scaling relations for sub-galactic regions of spatial scales ranging from $1\times1$\,kpc up to $4\times4$\,kpc.
For this comparison we used the SFR for each individual region (Table \ref{tab:models}) and the scaling relation for the corresponding spatial scale.
We have used the largest ($4 \times 4$\,kpc) fit for \textit{total} and the \textit{ring} that have even largest sizes. 
Region \textit{R2} shows an excess of $\rm \sim 0.4 \,dex \, (erg \, s^{-1} \, M_\odot^{-1} \, yr)$ with respect to all the models. 
The \textit{ring} shows lower $L_X$/SFR compared to \textit{R2} but still has an excess compared to the aforementioned models and empirical fits.

\begin{table*}
    \centering
    \caption{Properties of the sub-galactic regions defined in Figure\,\ref{fig:Maps} derived by optical data.}
    \setlength\tabcolsep{2.pt}
    \renewcommand{\arraystretch}{1.4}
    \begin{tabular}{c|c|c|c|c|c|l}
         Region & SFR (24$\mu$m) & SFR (12$\mu$m) & $M_\star$ & log sSFR & Metallicity & Age\\
         & $\rm M_\odot \, yr^{-1}$ & $\rm M_\odot \, yr^{-1}$ & $\rm 10^{10} \, M_\odot$ & $\rm M_\odot \, yr^{-1} / M_\odot$ & 12 + log(O/H) & Myr\\
         \hline
         \textit{total} & 8.60$\pm$0.30 & 6.11$\pm$0.04 & 2.78$\pm$0.01 & -9.51$\pm$0.02 & 8.39$\pm$0.17 & $5.2^{+9.0}_{-4.0}$\\
         \textit{bulge} & 3.20$\pm$0.18 & 1.77$\pm$0.02 & 0.64$\pm$0.01 & -9.30$\pm$0.02 & 8.50$\pm$0.04 & $6.7^{+54.0}_{-5.0}$\\
         \textit{inner bulge} & 0.99$\pm$0.10 & 0.34$\pm$0.01 & 0.11$\pm$0.01 & -9.02$\pm$0.04 & 8.50$\pm$0.04 & $7.2^{+56.2}_{-6.0}$\\
         \textit{bar} & 0.88$\pm$0.10 & 0.54$\pm$0.01 & 0.11$\pm$0.01 & -9.08$\pm$0.05 & 8.39$\pm$0.17 & $7.0^{+48.0}_{-4.5}$\\
         \textit{R2} & 0.58$\pm$0.08 & 0.22$\pm$0.01 & 0.04$\pm$0.01 & -8.88$\pm$0.06 & 8.13$\pm$0.03 & $6.0^{+8.0}_{-4.5}$\\
         \textit{ring} &  2.39$\pm$0.16 & 1.38$\pm$0.02 & 0.41$\pm$0.01 & -9.24$\pm$0.03 & 8.25$\pm$0.08 & $5.4^{+8.0}_{-4.0}$\\
    \end{tabular}
    \label{tab:models}
\end{table*}

\begin{figure}
    \centering
    \includegraphics[width=\columnwidth]{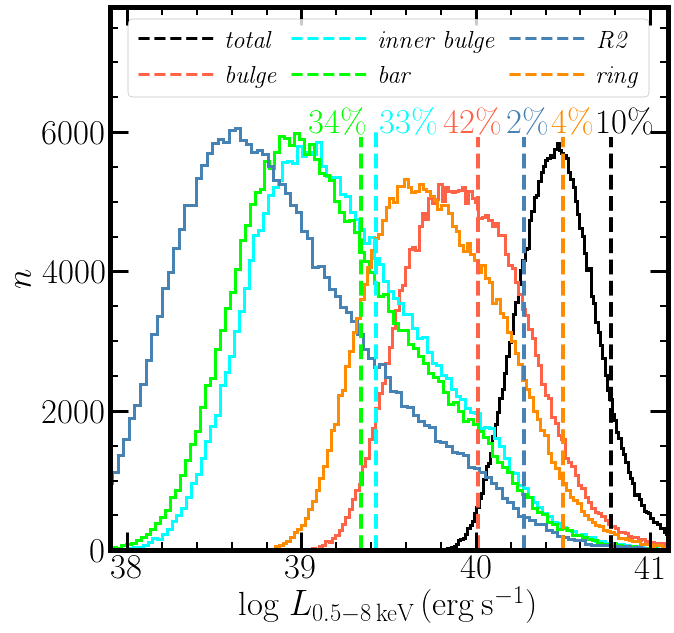}
    \caption{
    Expected X-ray luminosity (0.5--8\,keV) distributions of the different regions and the integrated emission of NGC\,922, drawn from the XRB XLF and the XRB scaling relations with stellar mass and SFR. 
    The vertical dashed lines with the same colour as the distributions indicate the measured X-ray luminosity of the corresponding region.
    The probability to have the measured luminosity (or higher), drawn out of these distributions is $10.27\%$, $42.31\%$, $33.40\%$, $34.33\%$, $1.56\%$, and $3.81\%$ for the \textit{total} emission and the regions of the \textit{bulge}, \textit{inner bulge}, \textit{bar}, \textit{R2}, and \textit{ring} respectively.
%    as shown at the top of the vertical lines. 
    }
    \label{fig:LX_XLF_dist}
\end{figure}

To assess whether this observed luminosity excess is a stochastic effect or it has a physical origin, we have followed a similar approach to \cite{2019MNRAS.483..711A}, where we have simulated the expected X-ray luminosity distribution in each sub-galactic region as well as the entire galaxy based on galaxy-wide XRB scaling relations with stellar mass and SFR (Fig. \ref{fig:LX_XLF_dist}).
In more detail,  we have calculated the total luminosity of the XRB populations by integrating the  XLF of low-mass XRBs \citep[LMXBs;][]{2004MNRAS.349..146G} and HMXBs \citep[][]{2012MNRAS.419.2095M} in each region above limiting luminosities of $\mathrm{\textit{L}_{min}=10^{36}erg\ s^{-1}}$  and  $\mathrm{\textit{L}_{min}=2\times 10^{37} erg\ s^{-1}}$  respectively.  The  expected number of LMXBs and HMXBs in each region (normalization of the XLF) was calculated using the scaling relations of \cite{2004MNRAS.349..146G} and \cite{2012MNRAS.419.2095M} respectively, and the local stellar mass and SFR (Table \ref{tab:models}).
To account for fluctuations on the number of sources, we have drawn 500 samples  from a Poisson distribution where its mean is equal to the expected number of LMXBs and HMXBs. 
To account also for stochastic effects on the luminosity of each region, we have obtained 500 samples of X-ray luminosity distributions from the corresponding 
XLF for each one of the 500 possible number of sources. 
This resulted in a distribution of 500,000 total XRB luminosities for each region, 250,000 originating from the LMXB and 250,000 from the HMXB population. 
Our results (Fig. \ref{fig:LX_XLF_dist}) show that indeed the high X-ray luminosities of the \textit{ring} and \textit{R2} have a very small probability to be the result of stochastic sampling (4\% and 2\% respectively).
This probability becomes 14\% for the \textit{ring} if we discard the  bright X-ray source associated with the \textit{R2} region, which contributes more than 50\% of the X-ray emission of the \textit{ring}.

\section{Discussion}
\label{sec:Discussion}

\subsection{Metallicity variations within the galaxy}

NGC\,922 shows overall near-solar to sub-solar metallicity ranging from $\rm8.11 \leq [12 + log(O/H)] \leq 8.67$. However, the most important finding is that regions on the ring have systematically lower metallicity than the bulge (Fig. \ref{fig:Metal}).  
\cite{2018MNRAS.476.5681E} found that NGC\,922 possesses a higher abundance of \ion{H}{I} gas (log\,$(M_{\rm \ion{H}{I}}/M_\star) = -0.50$) compared to normal galaxies of similar stellar mass (log\,$<M_{\rm \ion{H}{I}}/M_\star> = -0.89$).
This combined with the effect of the caustic that created the ring by displacing outwards the gas in the disc also explains the higher \ion{H}{I} mass-to-light ratio on the ring of NGC\,922 in comparison to its  bulge \citep[see Fig. 8 in][]{2018MNRAS.476.5681E}.

Several studies have suggested that the low gas phase metallicity of the RiG rings reflects the metallicity of the passing-through dwarf galaxy \citep[e.g.][]{1998AJ....116.2757B,2018MNRAS.481.2951E}.
This scheme is supported in NGC\,922, as we find that the ring shares similar metallicity ($\rm [12 + log(O/H)_{\textit{ring}}] = 8.25 \, \pm 0.08$) with the interacting companion 
($\rm [12 + log(O/H)] \simeq 8.3$; \citealt{{2006MNRAS.370.1607W}})
In this case the metallicity of the bulge is related with that of the host disk-like, more evolved galaxy.

The low metallicity of the ring regions with respect to the bulge region could also be the result of  the negative metallicity gradients typically seen in spiral galaxies \citep[e.g.][]{2010ApJS..190..233M}. 
If the star-forming activity is the result of the in-situ compression of gas of the original disk at the galactocentric radius of the ring, we would naturally expect that these star-forming regions will have lower metallicity than the bulge.
Alternatively, if the  ring consists of gas displaced from the inner parts of the galaxy, the negative metallicity gradient will result in the dilution of this gas with metal-poor material at the current location of the ring.
The latter scenario, however, would result into lower metallicity differences than the former.
This effect of course can be amplified by the lower metallicity of the intruder galaxy, which can also result in higher \ion{H}{I} abundance with respect to the stellar mass or star-light in the ring.

\subsection{Excitation of star-forming regions}

An interesting feature of the spectra from all regions is that they show \ion{He}{I} emission (Table \ref{tab:lines}), a strong indication that they host young stellar populations (age $<10$\,Myr). 
Furthermore, the ubiquitous presence of \ion{He}{I} and the short lifetimes of the stars with hot photospheres capable of exciting \ion{He}{I} suggests that their stellar populations have similar ages.

Indeed, \cite{2010AJ....139.1369P} found that both the ring and the bulge host very young star-clusters with ages $\simeq 7$\,Myr, with the bulge also hosting a population of older clusters ($30$--$350$\,Myr). 
More specifically, using the distribution of star-cluster ages of \cite{2010AJ....139.1369P}, we find that all the regions of interest in our analysis are dominated by young star clusters with ages around 5--7\,Myr (Fig. \ref{fig:SC_ages_dist}). 
The age distributions of the \textit{bulge} and \textit{inner bulge} show a tail to older clusters with a second, weaker, peak around 50\,Myr.
Clearly, the presence of extremely young star-clusters is consistent with our detection of \ion{He}{I} emission in all regions.

The location of the different regions on the BPT diagrams (Fig.\ref{fig:BPT}) can provide additional insights into their physical conditions.
All NGC\,922 regions examined here are encompassed by the \cite{2003MNRAS.346.1055K} curve, indicating purely star-formation driven ionization, without significant contribution from shock ionization (e.g. from supernovae and stellar winds in a young starburst). 
However, we see a clear segregation of the bulge and the ring regions in the $[$\ion{N}{II}$]$/\Ha\ BPT diagram:
The ring regions are located at the upper left of the \ion{H}{II}-region locus, while the bulge regions have lower $[$\ion{O}{III}$]$/\Hb\ ratios, and higher $\rm[\ion{N}{II}]$/\Ha\ ratios. 
Based on  the detection of \ion{He}{I} in both the ring and bulge regions, we interpret the different location of the regions on the BPT diagrams as the result of metallicity rather than age differences. 
Increasing metallicity (or decreasing ionization parameter) tends to move the locus of the points towards the right and lower part of the $[$\ion{N}{II}$]$/\Ha\ diagram, while it does not have as strong effect in the other diagrams \citep[e.g.][]{2001ApJ...556..121K,2006MNRAS.372..961K}.

The fact that we do not find any evidence of AGN activity in NGC\,922 is intriguing given the copious amounts of \ion{H}{I} gas in the galaxy and the expected presence of a super-massive black hole (SMBH), as in most galaxies \citep[e.g.][]{2013ARA&A..51..511K}.
This can be explained through three possible scenarios: 
a) AGNs are known to have a duty cycle \citep[e.g.][]{1966ApJ...146....7S,2005MNRAS.362...25B,2020ApJ...892...17D} and it is possible that currently the SMBH is in a low accretion state;
b) the gas has not lost its angular momentum yet and it has not reached the SMBH;
c) gravitational recoil of the SMBH due to the interaction may have displaced it out of the bulge. 
The last scenario is the most unlikely since we do not see a strong point like source outside the bulge (c.f. Fig. \ref{fig:Ha_c}).
In fact, the X-ray analysis in similar galaxies shows that an active AGN is not ubiquitous in RiGs \citep{2018ApJ...863...43W}.

\subsection{The X-ray emission of NGC922}

The X-ray spectra of the \textit{R2} and \textit{ring} regions can be well described solely by power-law emission with a photon index $\Gamma\sim1.7$. 
This is a strong indication that their spectra are dominated by XRBs emission.
The X-ray spectrum of the \textit{bulge} requires both a thermal ($\rm{kT=0.2}$\,keV) and a power-law  ($\Gamma \sim 1.9$) component similar to that of XRBs. 
The \textit{bar} region has a considerably softer X-ray spectrum, but consistent with the typical spectrum of XRBs ($1.7 \leq \Gamma \leq 2.5$). 
In the case of the \textit{inner bulge}, which is dominated by a thermal-plasma model, we can set a limit on the contribution of a power-law component.  These results are consistent with those of \citet{2012ApJ...747..150P}, who found a population of bright X-ray sources associated primarily with the ring and bar regions. 
The spectral parameters and total luminosity of the \textit{R2}, which hosts the brightest ULX in the galaxy, are consistent within the uncertainties with those reported in Section \ref{sec:X_ray_results}.  
No bright sources were found in the bulge.

We see a very similar pattern in the X-ray emission in the Cartwheel galaxy~\citep{2004A&A...426..787W}: diffuse thermal emission and a non-thermal component, described by a power-law model, due to the XRB population. The thermal plasma has a temperature of $\rm kT\,=0.2\,keV$, like in NGC922 with \nh\ = $\rm 2.3\times 10^{21} \, cm^{-2}$ for an $L_X^{\rm 0.5-10 \, keV} = 3 \times 10^{40} \, {\rm erg \, s^{-1}}$. 
The non-thermal emission can been divided in three different components: the brightest (hyperluminous) X-ray source (N10), the sum of detected point sources, and the residual non-thermal component due to the unresolved XRBs.
N10 is fitted by a power-law model with $\Gamma=1.6$, \nh\ = $\rm 3.6\times 10^{21} \, cm^{-2}$), similar to those of \textit{R2} in the NGC\,922 ring. The other point sources, both resolved and unresolved, have a steeper spectrum of $\Gamma=2.1-2.3$, \nh\ = $\rm 2\times 10^{21} \, cm^{-2}$. 
This is somewhat steeper than the spectrum of the NGC\,922 \textit{ring}, but consistent within the uncertainties. 
The most luminous source (N10) has a luminosity of  $L_X^{\rm 0.5-10 \, keV} = 1.4 \times 10^{41}$\erg\ dominating the X-ray emission of the galaxy, while the total additional contribution from all point sources is $L_X^{\rm 0.5-10 \, keV} = 1.2 \times 10^{41}$\erg. 
The most luminous source in NGC\,922 has a factor of $\sim5$ lower luminosity than the N10 source, and lower impact in its total luminosity.

\subsection{Effect of metallicity on X-ray luminosity of X-ray binary populations}

There is a growing body of observational evidence showing  strong anti-correlation between the number of luminous X-ray sources  and the metallicity of their host galaxies \citep[e.g.][]{2009MNRAS.395L..71M,2012ApJ...747..150P,2016MNRAS.457.4081B}. 
A similar trend holds for the integrated X-ray emission of galaxies, particularly those found at higher redshifts \citep[e.g.][]{2019ApJ...885...65F}. 
The clear metallicity difference between the bulge and the ring of  NGC\,922, combined with the information on the age of the stellar populations in these regions, provides an excellent test-bed for this X-ray luminosity--metallicity dependence.

Figure \ref{fig:LX_SFR} shows an anti-correlation between the $L_X$/SFR ratios and the metallicity of different regions in NGC\,922. Despite the large uncertainties (especially in the lower-luminosity -- higher metallicity regions) we see a systematic trend for lower metallicity regions to have stronger X-ray emission for their star-forming activity (i.e. higher $L_X$/SFR ratios). Although younger stellar populations may also result in elevated X-ray luminosities \citep[e.g.][]{2013ApJ...764...41F}, the observed relation is unlikely to be an age effect since, as discussed in the previous subsection, both the bulge and the ring regions host similarly young stellar populations (c.f. Table \ref{tab:models}). 

The observed relation shown in Figure\,\ref{fig:LX_SFR} agrees very well with the theoretical models of \citet{2013ApJ...776L..31F} and \citet{2017ApJ...840...39M}, and with the observational results of \citet{2016MNRAS.457.4081B} and \citet{2020MNRAS.495..771F}.  Although the $L_X$/SFR ratios of the higher metallicity bulge regions have large uncertainties, they appear to better follow the \cite{2017ApJ...840...39M} relation. 

While most regions are in good agreement with the theoretical and observational relations shown in Figure\,\ref{fig:LX_SFR}, region \textit{R2} lies above these relations, even after accounting for the observed scatter in the sub-galactic $L_X$--SFR scaling relations \citep[][]{2020MNRAS.494.5967K}. 
This region has the lowest metallicity ([12 + log(O/H)] = 8.13) and highest sSFR (log sSFR = $\rm -8.8 \, M_\odot \, yr^{-1}/M_\odot$) and it is $\sim 0.4$\,dex above any of these relations. 
This is because this region hosts a very luminous ULX ($L_X^{\rm 2-10 \, keV} \simeq 1.5 \times 10^{40} \, \rm erg \, s^{-1}$) which dominates its X-ray emission.
We do not expect the X-ray emission of such regions to follow the general $L_X$-SFR scaling relations since it is dominated by individual sources rather than the average XRB populations (in other words the central-limit theorem which is the basis for such relations does not hold). 
This is demonstrated in Figure\,\ref{fig:LX_XLF_dist}, which shows that the region \textit{R2} is highly inconsistent with these scaling relations even when we account for Poisson fluctuations of the number of XRBs and stochastic sampling of their XLF. 
Similarly, the slight excess of the \textit{ring} region with respect to expected $L_X$/SFR ratio for its metallicity is the result of the significant contribution of the ULX located in \textit{R2}. 
This source contributes $62\%$ of the 2--10\,keV X-ray luminosity of the \textit{ring} region. 
In fact, when we account for the contribution of the \textit{R2} region, the $L_X$/SFR ratio for the remaining ring agrees very well with that expected from the theoretical and observational relations. 

\subsection{Summary}
\label{sec:Summary}

In the previous sections we presented an analysis of the metallicity and excitation of star-forming regions in different regions of the ring galaxy NGC\,922 derived from optical data. We also analyzed the \cha\ X-ray data for the same regions in order to study the connection between metallicity and X-ray emission.  Our results can be summarized as follows:

\begin{enumerate}
    \item We observe a significant metallicity difference between the bulge ([12 + log(O/H)] $\sim 8.6$) and the ring ([12 + log(O/H)] $\sim 8.2$).
    \item All studied regions have systematically sub-solar metallicities with the bulge being marginally consistent with solar.
    \item We do not find any evidence for AGN activity in the bulge.
    \item We detect \ion{He}{I} emission in all regions indicating excitation from very young populations, supported by the typically less than $10$\,Myr ages of the star-cluster in the studied regions.  
    \item We observe an anti-correlation between the $L_X$/SFR and metallicity in NGC\,922. The similarity of the ages of the stellar populations in the studied regions suggests that this anti-correlation is primarily driven  by the effect of metallicity.
\end{enumerate}

\section*{Acknowledgements}

The authors thank the anonymous referee for constructive comments that helped to improve the clarity of the paper. 
We also thank Anne Pellerin for sharing the NGC\,922 star cluster  data, and Emanuele Ripamonti for obtaining the optical data used in this paper.
K. K., A. Z., and K. A. acknowledge funding from the European Research Council under the European Union's Seventh Framework Programme (FP/2007-2013)/ERC Grant Agreement n. 617001
(A-BINGOS). This project has received funding from the European Union's Horizon 2020 research and innovation programme under the
Marie Sklodowska-Curie RISE action, grant agreement No 691164 (ASTROSTAT). 
%AZ also acknowledges support from \textit{Chandra} grant G02-3111X.
AF acknowledges support by the Chandra X-ray Center (CXC) under National Aeronautics and Space Administration (NASA) contract NAS8-03060.
AW acknowledges financial support from ASI through the ASI-INAF agreements 2017-14-H.0.

This research has made use of: (a) observations collected at the European Southern Observatory under ESO programme 088.B-0882(A); (b) data obtained from the Chandra Data Archive and software provided by the CXC in the application packages DS9, CIAO, and Sherpa. The CXC is operated for NASA by the Smithsonian Astrophysical Observatory; (c) data products from the Wide-field Infrared Survey Explorer (WISE), which is a joint project of the University of California, Los Angeles, and JPL, California Institute of Technology, funded by NASA; (d) observations made with the Spitzer Space Telescope, which was operated by JPL, California Institute of Technology under a contract with NASA; (e) the NASA/IPAC Extragalactic Database (NED), which is operated by the Jet Propulsion Laboratory (JPL), California Institute of Technology, under contract with NASA; (f) the NASA/IPAC Infrared Science Archive (IRSA), which is funded by NASA and operated by the California Institute of Technology; (g) IRAF which was distributed by the National Optical Astronomy Observatory, which was managed by the Association of Universities for Research in Astronomy under a cooperative agreement with the National Science Foundation.

\section*{DATA AVAILABILITY}

The data underlying this article are available in the article and in its online supplementary material.
The raw data are available from the ESO and Chandra data archives.
%%%%%%%%%%%%%%%%%%%%%%%%%%%%%%%%%%%%%%%%%%%%%%%%%%

%%%%%%%%%%%%%%%%%%%% REFERENCES %%%%%%%%%%%%%%%%%%

% The best way to enter references is to use BibTeX:

\bibliographystyle{mnras}
\bibliography{main}

%%%%%%%%%%%%%%%%%%%%%%%%%%%%%%%%%%%%%%%%%%%%%%%%%%
%%%%%%%%%%%%%%%%% APPENDICES %%%%%%%%%%%%%%%%%%%%%

%\appendix

% Don't change these lines
\bsp	% typesetting comment
\label{lastpage}
\end{document}